\documentclass[11pt]{amsart}

\usepackage{amssymb}
\usepackage[pdftex]{graphicx}
\usepackage{hyperref}

\DeclareMathOperator{\Tr}{Tr}

\newcommand\tD{\tau_{\mathrm{d}}}
\newcommand{\rmd}{\mathrm{d}}
\newcommand{\rme}{\mathrm{e}}
\newcommand{\rmi}{\mathrm{i}}
\newcommand{\T}{\mathrm{T}}
\newcommand{\U}{\mathrm{U}}
\newcommand{\OO}{\mathrm{O}}
\newcommand{\St}{Z}
\newcommand{\Tt}{T}

\newcommand{\wb}[1]{{\overline{#1}}}
\newcommand{\wt}[1]{{\widetilde{#1}}}
\newcommand{\wh}[1]{{\widehat{#1}}}
\newcommand{\bs}[1]{{\boldsymbol{#1}}}

\newcommand{\V}{v}
\newcommand{\Vt}{\wt{v}}
\newcommand{\E}{e}

\newcommand{\Eu}{E_{\mathrm{u}}}
\newcommand{\Eo}{E_{\mathrm{o}}}

\newcommand{\Ni}{N_{i}}
\newcommand{\No}{N_{o}}

\newcommand{\fh}{\hat{f}}

\newtheorem{lemma}{Lemma}

\newtheorem{conjecture}{Conjecture}

\theoremstyle{definition}
\newtheorem{definition}{Definition}
\newtheorem{remark}{Remark}
\newtheorem{example}{Example}

\begin{document}

\title[Combinatorial theory of the semiclassical evaluation of transport moments II]
{Combinatorial theory of the semiclassical evaluation of transport moments II: \\
Algorithmic approach for moment generating functions}

\author{G.~Berkolaiko} 
\address{Department of Mathematics, Texas A\&M University, College
  Station, TX 77843-3368, USA}

\author{J.~Kuipers}
\address{Institut f\"ur Theoretische Physik, Universit\"at Regensburg, D-93040
Regensburg, Germany}

\begin{abstract}
  Electronic transport through chaotic quantum dots exhibits universal
  behaviour which can be understood through the semiclassical
  approximation.  Within the approximation, transport moments reduce
  to codifying classical correlations between scattering trajectories.
  These can be represented as ribbon graphs and we develop an
  algorithmic combinatorial method to generate all such graphs with a
  given genus. This provides an expansion of the linear transport
  moments for systems both with and without time reversal symmetry.
  The computational implementation is then able to progress several
  orders higher than previous semiclassical formulae as well as those
  derived from an asymptotic expansion of random matrix results.  The
  patterns observed also suggest a general form for the higher orders.
\end{abstract}

\maketitle

\section{Introduction}
\label{sec:intro}

The scattering matrix, which connects the asymptotic incoming and outgoing states, 
conceals the detailed scattering dynamics of the system.  Instead the entries encode 
the transport of states through or across the system.  We consider a cavity with 
two leads attached, so that the scattering matrix separates into reflection and 
transmission subblocks
\begin{equation}
\label{scatmatpartseqn}
S(E) = \left(\begin{array}{cc}\bs{r}&\bs{t}' \\ \bs{t} & \bs{r}'\end{array}\right). 
\end{equation}
If the leads carry $N_1$ and $N_2$ channels respectively, $\bs{r}$ is a square 
$N_1\times N_1$ matrix while $\bs{t}$ is $N_2\times N_1$.  We set $N=N_1+N_2$.  
The subblocks $\bs{r}$ and $\bs{t}$ encode the electronic transport from one lead 
to itself or the other respectively.  In particular, the eigenvalues of the 
matrix $\bs{t}^{\dagger}\bs{t}$ are the set of transmission probabilities whose sum is 
proportional to the average conductance through the cavity \cite{Buttiker86,Landauer57,Landauer88}.  

When the cavity is chaotic, the transport properties turn out to be
independent of the specifics of the system under consideration.  To
uncover this universal behavior, two methods have been developed: one
involves the semiclassical approximation for the scattering matrix
elements in terms of classical trajectories
\cite{miller75,richter00,rs02}, while the other is the random matrix
theory (RMT) approach of replacing the scattering matrix with a random
one chosen from the appropriate symmetry class
\cite{bm94,BluSmi_prl88,BluSmi_prl90,jpb94}.  The basic choice is whether the
system has time-reversal symmetry (TRS) or not; the corresponding
random matrices are the circular orthogonal ensemble (COE) and the
circular unitary ensemble (CUE) respectively.

Regardless of the method used, one of the characteristics of the transport properties are the linear moments
\begin{equation}
  \label{eq:def_linear_moment}
  M_n(X) = \left\langle \Tr \left[X^{\dagger}X\right]^{n} \right\rangle,
\end{equation}
where $X$ can be either the reflection $\bs{r}$ or transmission $\bs{t}$ subblock 
of the scattering matrix in \eqref{scatmatpartseqn}.  One can also consider ``non-linear'' moments, 
such the as the cross-correlation
\begin{equation}
  \label{eq:2_nonlinear}
  M_{n_1,n_2}(X_1, X_2) = \left\langle
    \Tr\left[X_1^{\dagger}X_1\right]^{n_1} 
    \Tr\left[X_2^{\dagger}X_2\right]^{n_2} 
  \right\rangle,
\end{equation}
where $X_1$ and $X_2$ are again certain sub-blocks of the scattering
matrix $S$.  Moments involving more products of traces can also be
considered.  In fact, on the ``other side'' of the spectrum, in a
sense, are the moments
\begin{equation}
  \label{eq:HCIZ_moment}
  M^k(X) = \left\langle \left(\Tr X^{\dagger}X\right)^{k} \right\rangle,
\end{equation}
whose generating function in the CUE case is a special case of the
celebrated Harish-Chandra--Itzykson--Zuber (HCIZ) integral
\cite{Har_ajm57,ItzZub_jpm80}
\begin{equation}
  \label{eq:HCIZ_integral}
  I_N(z, A, B) = \int_{U(N)} e^{-zN \Tr(A U B U^{\dagger})} \rmd U.
\end{equation}
Typically one sets $X$ to be the transmission
subblock $\bs{t}$ of the scattering matrix in \eqref{scatmatpartseqn}
so that \eqref{eq:def_linear_moment} becomes the moments of the
transmission eigenvalues and \eqref{eq:HCIZ_moment} becomes the
moments of the conductance.  

In the companion paper \cite{bk13a}, we showed that the RMT and
semiclassical results for any of these moments must be identical.
However, while the equivalence has been established, the problem of
calculating particular moments is still not fully answered, and
remains a hard challenge.  There is also a wider class of problems
related to energy differentials of the scattering matrix, to
systems with superconducting leads attached, and to systems with
non-zero Ehrenfest time which can and have also been considered, both
semiclassically and within RMT.  We review these results in
Section~\ref{sec:transmoms}.

\begin{figure}[t]
  \includegraphics[scale=1]{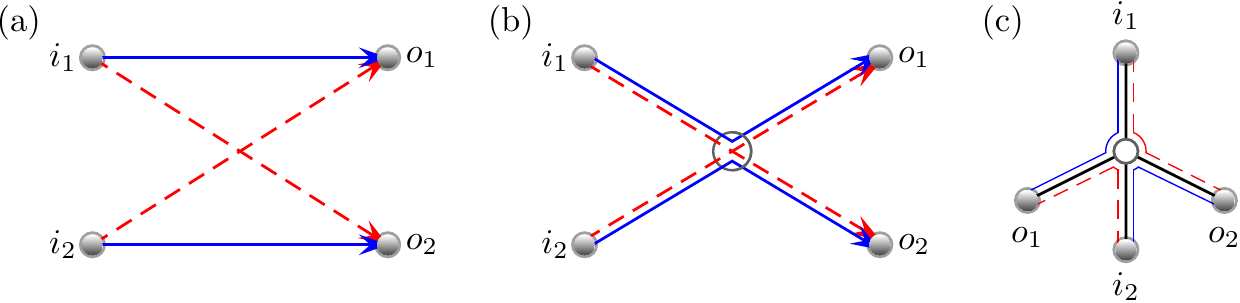}
  \caption{(a) A quadruplet of trajectories that appear in the second
    linear moment.  (b) For the actions to (nearly) cancel, the blue
    solid and the red dashed trajectories must coincide pairwise along
    the most part of their length.  Nontrivial (but significant)
    contributions arise when pairs exchange partners by coming close
    to each other (``crossing'') in encounter regions.  (c) A ribbon
    graph representation of the quadruplet where the encounter becomes
    a roundabout vertex, links become edges and the trajectories
    create a boundary walk.}
  \label{fig:secondmomentexample}
\end{figure}

Here we focus mostly on the linear moments in \eqref{eq:def_linear_moment}, 
into which we substitute the semiclassical approximation \cite{miller75,richter00,rs02}
\begin{equation} 
  \label{scatmatsemieqn}
  S_{oi}(E) \approx \frac{1}{\sqrt{N\tD}}\sum_{\gamma (i \to o)}
  A_{\gamma}(E)\rme^{\frac{\rmi}{\hbar}S_{\gamma}(E)} ,
\end{equation}
involving the scattering trajectories $\gamma$.  These trajectories 
travel from incoming channel $i$ to outgoing channel $o$ with
action $S_{\gamma}$ and stability amplitude $A_{\gamma}$ (incorporating the Maslov phase) 
while $\tD$ is the average time spent inside the cavity.  The linear moments become the sum
\begin{equation}
  \label{semitrajeqn}
  M_n(X) \sim \left\langle \frac{1}{{(N\tD)}^{n}}
    \sum_{{i_j,o_j}} 
    \sum_{\substack{\gamma_j(i_j\to o_j) \cr
        \gamma'_j (i_{j+1}\to o_{j})}} 
    \prod_{j=1}^{n} A_{\gamma_j}A_{\gamma'_j}^{*}
    \rme^{\frac{\rmi}{\hbar}(S_{\gamma_j}-S_{\gamma'_j})} \right\rangle,
\end{equation}
where $i_{n+1}=i_1$ which endows the set of trajectories with a particular structure.  
Namely, moving forwards along the trajectories $\gamma_j$ and backwards along 
$\gamma'_j$ we visit the channels $i_1\to o_1 \to i_2 \ldots o_{n} \to i_1$ along 
a single cycle.  For $n=2$ this is depicted in Fig.~\ref{fig:secondmomentexample}(a).

However, to contribute consistently in the semiclassical limit of
$\hbar\to0$, the total action of the set of trajectories should be
stationary (under some average).  This can be achieved by forcing the
trajectories to be nearly identical, except in small regions known as
encounters.  An example for $n=2$ is given in
Fig.~\ref{fig:secondmomentexample}(b), with the encounter region
denoted by the unfilled circle.  As detailed in \cite{bk13a} and
explained in Section~\ref{sec:diagrams_for_linmoments} below, the
resulting diagram can be interpreted as a ribbon graph, depicted in
Fig.~\ref{fig:secondmomentexample}(c).  Edges of the graph correspond
to semiclassically long stretches of trajectories following each other
in pairs.  Vertices of degree $>1$ (``internal vertices'') correspond
to encounter regions where two or more pairs of orbits exchange
partners.  Vertices of degree $1$ (``leaves'') correspond to a pair of
trajectories entering or leaving the cavity via a lead.

What is particularly important
is that the semiclassical contribution of a diagram can easily be read
off from its structure \cite{heusleretal06,jw06,mulleretal07,wj06}:

\begin{definition}
  \label{def:Essen_ansatz}
  The semiclassical contribution of a diagram is a product where
  \begin{itemize}
    \item every edge provides a factor $1/N$,
    \item every internal vertex gives a factor of $-N$, 
    \item encounters that happen in the lead do not count (give a factor
      of $1$).
  \end{itemize}
\end{definition}

The diagram in Fig.~\ref{fig:secondmomentexample}(b) or (c) then gives the contribution 
\begin{equation*}
\sum_{\substack{i_1,i_2 \cr o_1,o_2}} \frac{-N}{N^4} ,
\end{equation*}
where the result for each channel sum is simply the number of channels in the respective lead.  
For example, when $X=\bs{t}$ the result is $-N_1^{2}N_2^{2}/N^3$, and
when $X=\bs{r}$, the result is $-N_1^4/N^3$.

With these simple rules, the task of evaluating transport moments
semiclassically reduces to that of systematically generating all
permissible diagrams.  This is itself a formidable problem, and
previous incremental progress in its solution is reviewed in
Section~\ref{sec:transmoms}.  In this manuscript we present an
algorithm that in principle allows one to calculate the generating
function 
\begin{equation*}
  \sum_{n=1}^\infty s^n \Tr \left[X^{\dagger}X\right]^{n},
\end{equation*} 
to any required order in the small parameter $1/N$.  We stress that
the answers obtained are for moments of \emph{all} orders $n$ at once.

To describe the algorithm we will seek a more detailed understanding of
the structure of the semiclassical diagrams contributing at a given
order and describe an algebraic method to generate them.  The
resulting algorithm is implemented on a computer, resulting in an
expansion that goes several orders beyond the best of the previously
available results \cite{bk11}.  In principle, the algorithm is
applicable to any order of $1/N$, but in practice it is severely
limited by the available computer capacity.

To classify the contributing orbits we go through several steps.
First, after explaining the structure of semiclassical diagrams, we
incorporate the contributions of diagrams with encounters in the leads
\cite{wj06} into the contribution of the ``principal diagrams'' (this
method goes back to \cite{bhn08}).  This is done in
Section~\ref{sec:diagrams_for_linmoments}.  Then, in
Section~\ref{sec:basestructures}, we argue that one can obtain
diagrams for arbitrary $n$ but fixed order of $1/N$ by grafting a
number of trees on the edges and vertices of a ``base structure''.
Most importantly, the number of such structures is finite for any
given order of $1/N$; the structures can be generated automatically
from factorizations of permutations as detailed in
section~\ref{sec:base}.  In Section~\ref{sec:algoapproach} we obtain
the semiclassical contributions of vertices and edges of a \emph{base
  structure}; such contributions are essentially the result of a
partial sum over all possible trees that can be grafted on the vertex
or edge.  In Section~\ref{sec:calculation} we present the
expressions for the moment generating functions resulting after
summation over all base structures at orders $N$ to $N^{-3}$.
Conjectures for higher orders are also given.

\section{Transport moments}
\label{sec:transmoms}

Before turning to a semiclassical method to obtain explicit results for 
transport moments, we first review some of the previous results in
this direction. The body of literature on RMT is immense due to its diverse
applications, from number theory to high energy physics (see
\cite{RMT_handbook} for a collection of articles reviewing properties
and applications of RMT).  Here we aim to review the particular
results that relate to our main task, to understand the linear moments
$M_n$ in \eqref{eq:def_linear_moment}, for large $n$.

\subsection{Previous RMT results}
\label{sec:previousrmt}

The first RMT approaches considered correlators of arbitrary products of matrix elements, 
which include all the types of moments discussed in the introduction.  Averaging over 
the CUE or the COE, results were obtained \cite{bb96,Mel_jpa90,Sam_jmp80} in terms of class 
coefficients or ``Weingarten'' functions.  Although they can in principle be used to calculate 
any moment, the class coefficients are generated recursively and the results become more
unwieldy as the order of the moments increases.  The problem became one of finding closed 
form results for higher moments.

To proceed, Brouwer and Beenakker \cite{bb96} developed a diagrammatic
approach to the random matrix integrals which, aside from recreating
previously known results for the conductance and its variance
\cite{bm94,jpb94}, allowed them to obtain the probability distribution
(and hence indirectly all the moments) of the transmission eigenvalues
at leading and subleading order in inverse channel number.  This
diagrammatic approach could also be applied to obtain various terms
when the scattering matrix is coupled to the leads via a tunnel
barrier, or for a normal metal-superconductor junction.

In order to obtain high moments beyond a diagrammatic expansion, 
a different approach was pioneered by Savin and Sommers \cite{ss06}.  Starting from the
probability distribution of the transmission eigenvalues of the matrix
$\bs{t}^{\dagger}\bs{t}$ \cite{beenakker97,forrester06} they noted the
similarities to the Selberg integral.  This allowed them to obtain the
second linear moment \cite{ss06} (related to the shot noise power) and
later all linear and non-linear moments up to fourth order
\cite{ssw08}.  Although moments of this order could still be tractable
using the recursive class coefficients, this work spurred a renewal of 
interest in the RMT treatment of transport moments.  For systems without 
TRS a result for all the linear moments as well as all the moments of the conductance,
\eqref{eq:HCIZ_moment}, were obtained using generalisations of the
Selberg integral in \cite{novaes08}.

The joint probability distribution of the transmission eigenvalues is a
particular case of the Jacobi ensemble in RMT \cite{forrester06} which
also allowed the linear moments to be calculated for the unitary case
using orthogonal polynomial techniques \cite{vv08} (results using a
variation of the Selberg integral were also obtained).  The linear
moments for all the classical symmetry classes as well as for the
superconducting symmetry classes were likewise obtained
using orthogonal polynomials \cite{lv11,ms11}.  These techniques were
also applied to the Laguerre ensemble and the linear moments of the
Wigner delay times were calculated \cite{ms11}.

For moments of the type \eqref{eq:HCIZ_moment}, a connection to the
theory of integrable systems was exploited to calculate all the
moments of the conductance \cite{ok08} and later the shot noise
\cite{ok09} for systems with broken TRS.  The same results, plus the
moments for systems with TRS, were obtained using generalised Selberg
integrals and symmetric functions \cite{kss09}.  The integrable system
approach for the moments of the conductance and shot noise was
recently extended to all the classical and superconducting symmetry
classes as well as for the moments of the Wigner delay time
\cite{ms13}.

Returning to the linear moments of the transmission eigenvalues, in
the case of broken TRS there exist several different expressions
\cite{lv11,ms11,novaes08,vv08}, each involving sums over
combinatorial-type terms.  The number of terms in the sums increase
with the order of the moments leaving high moments difficult to
obtain.  Interestingly, due to being obtained by different methods,
all the results look remarkably different despite encoding the same
object.  The asymptotic analysis in the limit of a large number of
channels $N$ is also challenging, especially beyond the leading order.
However, the results of \cite{ms11} which include systems with TRS are
more amenable for such an asymptotic expansion, as detailed in
\cite{ms12}.

\subsection{Previous semiclassical results}
\label{sec:previoussemiclassics}

Similar to RMT, on the semiclassical side the low moments were
obtained first starting with the conductance
\cite{heusleretal06,rs02}, the shot noise
\cite{braunetal06,SchPuhGei_prl03,wj06} and then the conductance variance
as well as other second order correlation functions
\cite{mulleretal07}.  Interestingly, it was the simple result for the
shot noise \cite{braunetal06} which had not yet been explicitly
calculated using RMT which prompted Savin and Sommers to revisit the
RMT approach \cite{ss06}.  All these semiclassical results were
obtained by mapping the semiclassical diagrams for open systems to
those which contribute to the spectral statistics of closed
quantum chaotic systems \cite{mulleretal04,mulleretal05,sr01}.  As the
order of the moment increases, this mapping becomes much more
complicated, although recently Novaes succeeded in relating this mapping to
various combinatorial problems whose solution allows the moments to be
generated for systems with broken TRS \cite{novaes12,novaes13}.

Taking a different approach \cite{bk12,bk13a} we could show that the
contribution of the vast majority of semiclassical diagrams cancel and
the remaining diagrams could be identified with primitive
factorisations of a permutation.  These give results identical to the
computation via the class coefficients (Weingarten functions) used for
matrix element correlators in RMT \cite{bb96,Mel_jpa90,Sam_jmp80} and
complete equivalence was thus established.  This approach works for
systems both with and without TRS but provides no further results for
the moments.  In a separate development, Novaes recently announced a
way to generate the diagrams with broken TRS from a matrix integral
which he also shows to give the RMT results for arbitrary moments and
prove the complete equivalence of semiclassics and RMT
\cite{novaes13b}.

In the quest for computing actual answers, most recent progress came
by looking in the direction of high moments but only to the first few
terms in the $1/N$ expansion.  First it was noticed that the
semiclassical diagrams which contribute at leading order to the linear
moments could be reinterpreted as trees \cite{bhn08,bhn08b} allowing
the moments of the transmission eigenvalues to be generated
recursively and encoded in a moment generating function \cite{bhn08}.
Including an energy dependence in the semiclassical contributions then
allowed access to the leading order density of states of Andreev
billiards \cite{kuipersetal11,kuipersetal10}.  Remarkably, for
transport through Andreev dots \cite{wj09}, the effect of the
superconducting leads means that complete tree recursions are
necessary even for the leading order contribution to the conductance
\cite{ekr11} (i.e.\ for calculating a low moment).  Energy dependent
correlation functions can also be related to the moments of the Wigner
delay times and the leading order moment generating function
correspondingly obtained \cite{bk10}.  Building on the semiclassical
treatment for low moments with tunnel barriers
\cite{kuipers09,whitney07}, the corresponding leading order generating
functions for the transport quantities and the moments of the
reflection eigenvalues in Andreev billiards were obtained in
\cite{kr13}.

These leading order results all agreed with the corresponding results
obtained by RMT (whenever the RMT answers were available)
\cite{bmb95,bb96,bfb99,melsenetal96,melsenetal97,sss01}.  However,
including an energy dependence or tunnel barriers changes the
semiclassical contributions in Definition~\ref{def:Essen_ansatz}.
Semiclassical diagrams no longer cancel each other completely, so the
proof of the equivalence of semiclassics and RMT \cite{bk12,bk13a} no
longer holds.  Of these other physical situations mentioned above, it
is only for the moments of the Wigner delay times though that a
general RMT result is known \cite{lv11,ms11,ms13} and where a proof of
the equivalence between semiclassics and RMT would currently be
feasible.  A proof of this, and progress on both sides for the other
cases would therefore be welcome.  As a further physical example, RMT
results are also known for the superconducting ensembles
\cite{ms11,ms13}, though the results are not in the form of the
scattering matrix correlators that were used in \cite{bk13a}.  This
suggests that a mapping from the semiclassical diagrams through
combinatorial objects to the RMT results may be significantly more
complicated than for the standard symmetry classes.

Beyond the remit of RMT, the semiclassical approach can handle the
effect of the Ehrenfest time, or the time over which an initially
localised quantum wavepacket spreads to the system size, which has
been studied for lower moments
\cite{Ada_prb03,br06,br06b,jw06,petitjeanetal09,RahBro_prl06,wk10,waltneretal12,whitney07,wj06}.
A result has also been obtained for all the linear moments at leading
order \cite{wkr11} which in particular leads to interesting signatures
in the density of states of Andreev billiards
\cite{kuipersetal11,kuipersetal10}.

Beyond leading order, a method for generating all semiclassical
diagrams at a particular order was developed in \cite{bk11}.  In
particular, moment generating functions were obtained up to second
subleading order for a range of transport moments.  Later, the
asymptotic expansion \cite{ms12} of the RMT results for the linear
moments of the transmission eigenvalues and the Wigner delay times
\cite{ms11} could recreate those generating functions.  The method in
\cite{bk11} becomes unwieldy for further subleading orders and so in
Section~\ref{sec:algoapproach} we develop a more powerful algebraic
approach which can likewise be used to compute the moments of the
Wigner delay times and the density of states of Andreev billiards.  In
Section~\ref{sec:calculation} it is applied to the calculation of
linear moments $M_n(X)$ for all values of $n$.  The answers are
obtained in the form of generating functions with respect to $n$,
asymptotically as $N\to\infty$ under the assumption that the size
$\Ni\times \No$ of the subblock $X^{\T}$ grows proportionally to $N$.
As a first step towards these results we organize the semiclassical
diagrams to allow for their efficient generation.

\section{Diagrams for the linear moments}
\label{sec:diagrams_for_linmoments}

We now turn to a combinatorial interpretation of the semiclassical
diagrams for the linear moments.  

\subsection{Semiclassical diagrams}

As outlined in Section~\ref{sec:intro}, an $2n$-tuple of trajectories
$\{\gamma_j, \gamma_j'\}$ contributes consistently in the
semiclassical limit if any given $\gamma_j'$ runs along some parts
of the trajectories $\{\gamma_j\}$ at all times, sometimes switching
from following one $\gamma$-trajectory to another.  For the switching
to happen, the two $\gamma$-trajectories have to come close in phase 
space.  The (semiclassically small) region where the switching
occurs is called the \emph{encounter region}.

\begin{figure}[t]
  \includegraphics[scale=0.7]{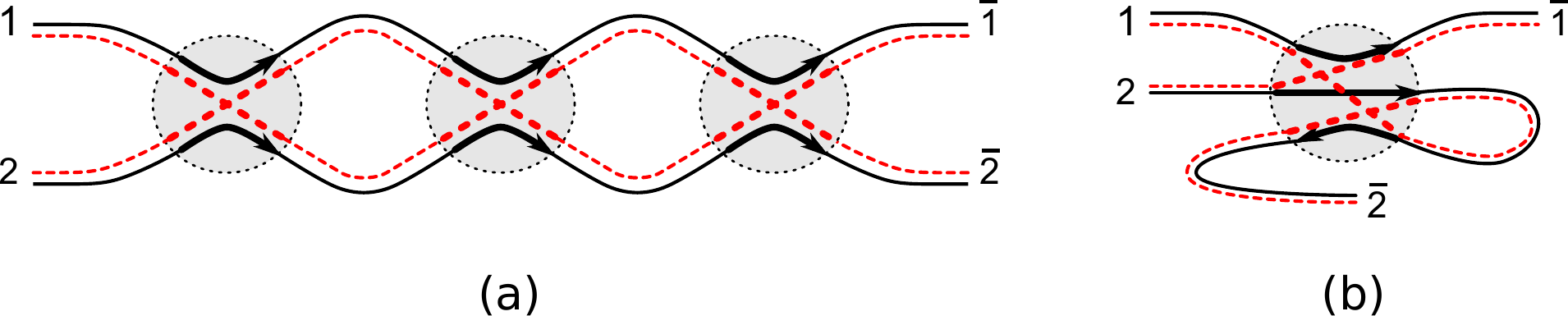}
  \caption{Two examples of semiclassical diagrams as drawn in the
    physics literature.  These examples correspond to diagrams (d) and
    (e) of Fig.~4 in \cite{mulleretal07}}
  \label{fig:encounter_physics_prin}
\end{figure}

A \emph{semiclassical diagram} is a schematic depiction of the
topology of the $2n$-tuple $\{\gamma_j, \gamma_j'\}$.  It describes
which part of the trajectory $\gamma_j'$ runs along which part of the
trajectory $\gamma_k$ and what gets switched with what in the
encounter region.  Examples of diagrams typical in the physics 
literature are shown in Fig.~\ref{fig:encounter_physics_prin}.
Note that to avoid clutter we often shorten labels $i_j$ to $j$ and
labels $o_j$ to $\wb{j}$.  In Fig.~\ref{fig:encounter_physics_prin}
the trajectories $\gamma_1$ and $\gamma_2$ running from $1$ to
$\wb{1}$, and from $2$ to $\wb{2}$ correspondingly, are shown as solid
black lines.  The trajectories $\gamma'$ are shown as dashed lines,
while encounter regions are shown as shaded circles.  In
Fig.~\ref{fig:encounter_physics_prin}(a), the trajectory $\gamma_1'$
(running from $2$ to $\wb{1}$) runs first along $\gamma_2$, then along
$\gamma_1$, then $\gamma_2$ and finally $\gamma_1$ again.  In
Fig.~\ref{fig:encounter_physics_prin}(b), trajectory $\gamma_2'$
starts from $1$ along $\gamma_1$, then follows $\gamma_2$ in the
direction \emph{opposite} to the direction of $\gamma_2$, finally
switching to another part of $\gamma_2$, now in the same direction.
This diagram requires TRS to contribute.

\begin{figure}[t]
  \includegraphics[scale=0.7]{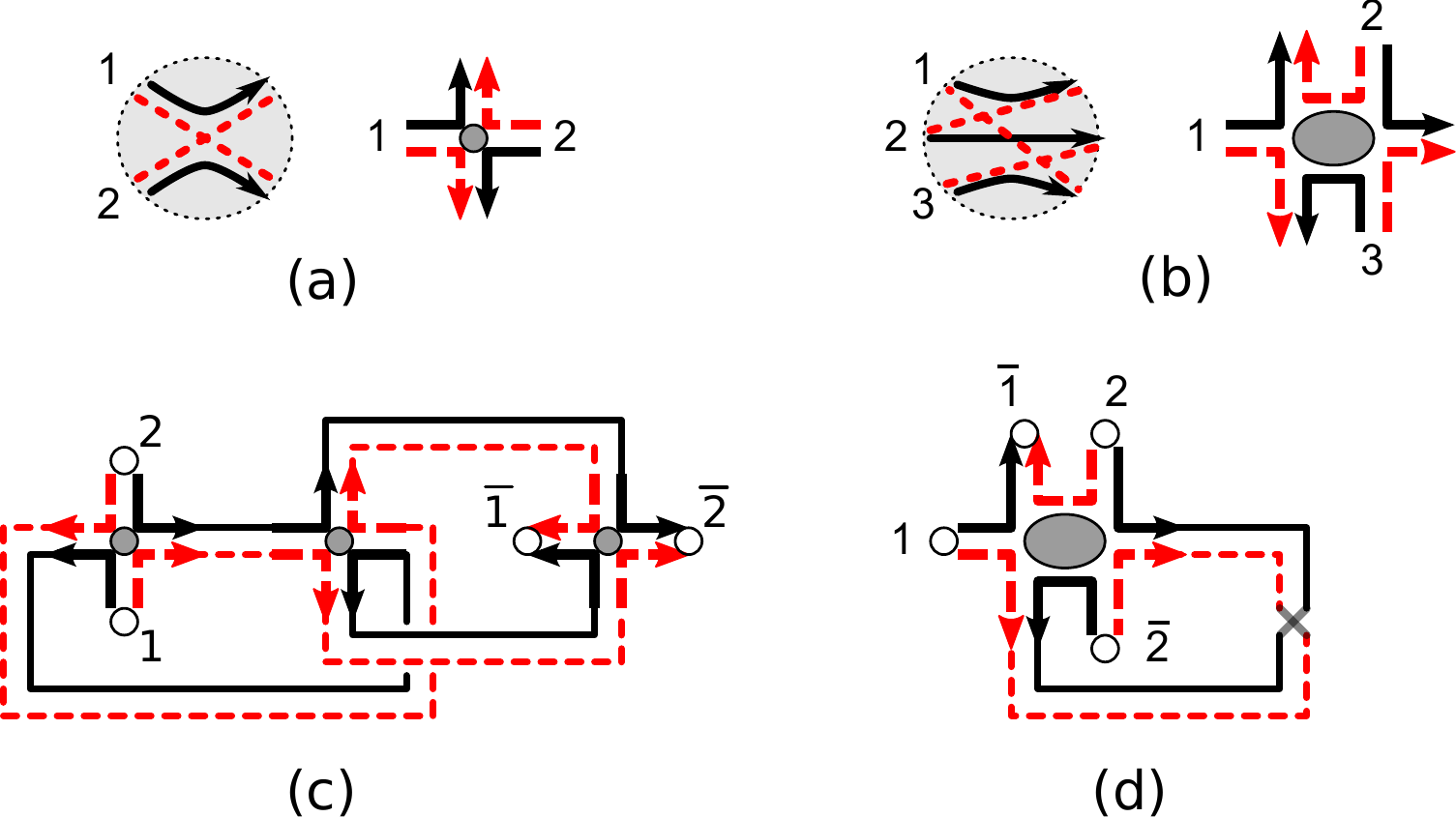}
  \caption{Untwisting the encounters into the vertices of the ribbon
    graph, (a) and (b).  The ribbon graphs (c) and (d) correspond to the
    diagrams of Fig.~\ref{fig:encounter_physics_prin}.  To
    read off the trajectories, we start at the open end labelled $1$
    and follow the left side for $\gamma$ or the right side for
    $\gamma'$.  The leaves (vertices of degree 1) of the graph are
    shown as empty circles; the internal vertices are represented by
    the filled ellipses.  Edges going to leaves are normally drawn
    short to save space.  Other edges often have rectangular
    corners; the corners carry no particular meaning and were only employed 
    due to the lack of artistic skill.}
  \label{fig:encounter_untwist_prin}
\end{figure}

Starting with \cite{bhn08} and especially in \cite{bk11}, it was
realized that ``untwisting'' the encounter region so that trajectories
do not intersect (see Fig.~\ref{fig:encounter_untwist_prin}), one
obtains an equivalent picture but with a significant advantage: it is
an object well studied in combinatorics and in some RMT literature, a
(combinatorial) map.  This term refers to a graph that is drawn on a
surface without self-intersections.  An important consequence of being
drawn is that the ordering of edges around every vertex is fixed.  If
one traces a path along one side of an edge, upon arrival at a vertex
there is a unique choice of the edge and the side along which to
continue.  This defines the boundary of the map.  If the boundary is
connected, the map is called \emph{unicellular}.  In-depth information
about maps can be found, for example, in
\cite{JacVis_Atlas,Tutte_GraphTheory}; the reader is referred to
\cite{Zvo_mcm97} for an especially accessible introduction with
applications to RMT.

To highlight the boundary of a map, the edges are often thickened in a
drawing (hence the alternative names ``ribbon graph'' or ``fat
graph'').  This is the approach we take.  The vertices of our maps are
drawn as circles (or ellipses).  Vertices of degree more than one are
shaded, they correspond to the encounters.  Vertices of degree one are
unfilled, they are henceforth called \emph{leaves} and correspond to
the initial or final points of the trajectories.  The edges of the map
are shown as parallel curves connecting the vertices.  The edges can
have right angle turns in them (due to our lack of drawing skill) and
M\"obius-like twists.  The latter are essential features of a map and
indicate that the map can only be drawn on a non-orientable surface
and the corresponding diagram requires TRS to contribute.  The
trajectories can now be read off as the sections of the boundary going
from one leaf to another.  As before, trajectories $\gamma_j$ are
drawn in solid lines, while $\gamma_j'$ are drawn dashed.  The
differences between unitary diagrams (with broken TRS) and orthogonal
diagrams (contributing in the presence of TRS) and some other features
are discussed after we introduce the \emph{principal diagrams} in the
next section.

\subsection{Principal diagrams and untying}

\begin{figure}[t]
  \centering
  \includegraphics[scale=0.6]{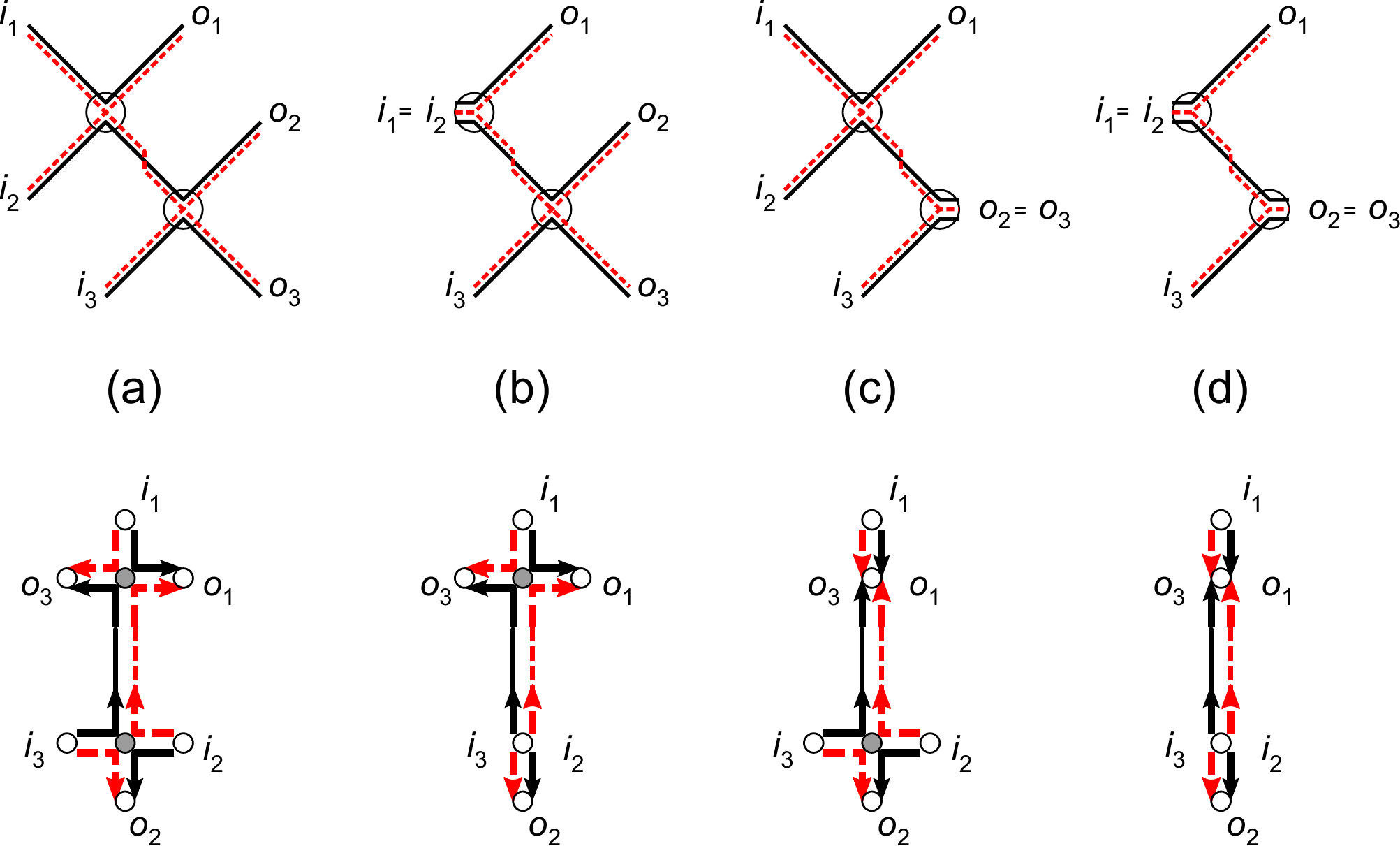}
  \caption{An example of a principal diagram and its untied versions
    contributing to the third moment.}
  \label{fig:third_moment_example}
\end{figure}

An example of a diagram which contributes to the third moment is depicted in 
Fig.~\ref{fig:third_moment_example}(a).  This diagram has two
encounters that happen inside the cavity, and, according to the rules
in Definition~\ref{def:Essen_ansatz}, its contribution is $(-N)^2/N^7$
(multiplied by $N_1^3N_2^3$ in the transmission case, once the
summation over all possible incoming and outgoing channels is
performed).  However, there is a related diagram obtained by moving
the first encounter close to the incoming lead, see
Fig.~\ref{fig:third_moment_example}(b).  From geometric
constraints, it follows that the channels $i_1$ and $i_2$ must
coincide for this to be possible.  According to
Definition~\ref{def:Essen_ansatz}, this diagram has a different
contribution.  Indeed, the two edges leading to the encounter
disappear and the encounter itself does not contribute anything.  The
resulting contribution is $N_1^2N_2^3 (-N)/N^5$.  Note the reduced
power of $N_1$ due to the summation restricted by $i_1=i_2$.
Importantly, when $N_1\sim N_2 \sim N$, the overall order of the
contribution does not change.

Similarly, one can move the right encounter close to the outgoing
lead, Fig.~\ref{fig:third_moment_example}(c) or move both encounters,
Fig.~\ref{fig:third_moment_example}(d).  The corresponding
semiclassical contributions to the moment are $N_1^3 N_2^2 (-N)/N^5$
and $N_1^2N_2^2 /N^3$.  On the lower half of
Fig.~\ref{fig:third_moment_example} the same diagrams are drawn as
combinatorial maps.  Note that in the map of
Fig.~\ref{fig:third_moment_example}(b) the lower vertex can be viewed
as two vertices (each of degree one) shown on top of each other; the
diagram then falls into two connected components.  Thus an encounter
happening in the lead is represented by ``untying'' the corresponding
encounter in the diagram.

It is instructive to explore the parallels between the encounter
happening in the lead (or untying a vertex in a diagram) with an
expansion of the corresponding moment in random matrix theory.

\begin{example}
  \label{exm:organization1}
  Consider the RMT result for the second moment in the unitary case without TRS.
  We are going to use the general formula
  \begin{equation}
    \label{eq:avU}
    \langle U_{a_1a_\wb{1}}\ldots U_{a_sa_\wb{s}} 
    U^*_{b_1b_\wb{1}}\ldots U^*_{b_tb_\wb{t}} \rangle_{\mathrm{CUE}(N)} = 
    \delta_{t,s} \sum_{\sigma,\pi\in S_t} V^{\U}_N(\sigma^{-1}\pi)
    \prod_{k=1}^{t} \delta\!\left({a_k}-{b_{\sigma(k)}}\right) 
    \, \delta\!\left({a_\wb{k}}-{b_\wb{\pi(k)}}\right),
  \end{equation}
  where $S_t$ is the symmetric group of permutations of the set
  $\{1,\ldots, t\}$, $\delta_{k,n} = \delta(k-n)$ is the Kronecker
  delta (the latter notation is used solely to avoid nesting
  sub-indices) and the coefficient $V^{\U}_N(\sigma^{-1}\pi)$ depends
  only on the lengths of cycles in the cycle expansion of
  $\sigma^{-1}\pi$, i.e.\ on the conjugacy class of the permutation
  $\sigma^{-1}\pi$.  This formula and the class coefficients
  $V^{\U}_N$ were first explored in detail by Samuel \cite{Sam_jmp80},
  although recently $V^{\U}_N$ became known as the ``unitary
  Weingarten function'' (after  \cite{Wei_jmp78}).

  Applying this formula to the second moment, setting $\St = S^{\T}$ we get
  \begin{align}
    M_2(X) &= \left\langle \sum_{\substack{i_1,i_2\cr o_1,o_2}} 
    \St_{i_1,o_1} \St_{i_2,o_2} \St^*_{i_2,o_1} \St^*_{i_1,o_2} \right\rangle
    \nonumber \\
    \label{eq:V_with_delta0}
    &=\sum_{\substack{i_1,i_2\cr o_1,o_2}}  \left[ V^{\U}(\tau) 
      + \delta_{i_1,i_2} V^{\U}( (1\, 2)\tau)
      + \delta_{o_1,o_2} V^{\U}( \tau(1\, 2)) 
      + \delta_{i_1,i_2}\delta_{o_1,o_2} V^{\U}( (1\, 2)\tau (1\, 2) )
    \right] \\
    \label{eq:V_with_delta}
    &= \Ni^2 \No^2 V^{\U}(\tau) + \Ni \No^2 V^{\U}( (1\, 2)\tau)
    + \Ni^2 \No V^{\U}( \tau(1\, 2)) + \Ni \No V^{\U}( (1\, 2)\tau (1\, 2) ).
  \end{align}
  Here $\Ni \times \No$ is the size of the subblock $X^{\T}$ and $\tau
  = (1\, 2)$ is called the \emph{principal target permutation}, given
  by $\tau = \sigma^{-1}\pi$, where the permutations $\sigma = (1\,
  2)$ and $\pi = id$ map the first and last indices of $\St$ to the
  first and last indices of $\St^*$.  This choice of $\sigma$ and
  $\pi$ is the only one available if the channels $i_1$, $i_2$ and
  $o_1$, $o_2$ are distinct.  If $i_1=i_2$, there is an additional
  possibility $\sigma=id$ accounted for by the second term in
  \eqref{eq:V_with_delta} and so on.

  The arguments of the functions $V^{\U}$ are formatted to highlight
  the connection to untying the diagrams.  Multiplication of the
  permutation $\tau$ by $(1\,2)$ on the left corresponds to untying
  the ends $i_1$ and $i_2$ of a diagram.  Multiplication by $(1\,2)$
  on the right is the untying of the ends $o_1$ and $o_2$.  This
  combinatorial encoding of untyings is explored in-depth in the
  Appendix.  We remind the reader that we often shorten the leaf
  labels $i_j$ to $j$ and $o_j$ to $\wb{j}$.
\end{example}

\begin{figure}[t]
  \includegraphics[scale=0.8]{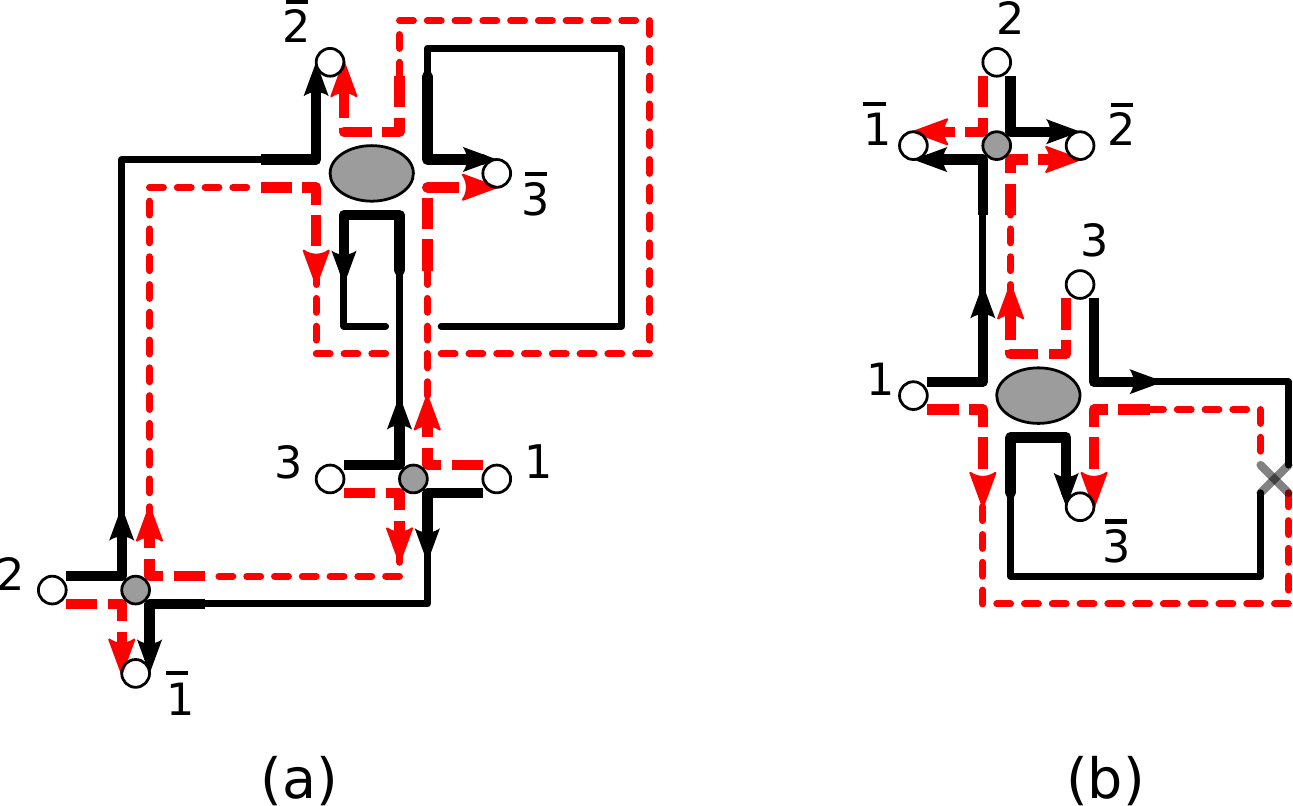}
  \caption{An example of a unitary and an orthogonal principal
    diagrams.  The shaded circles (and ellipses) represent the
    vertices of even degrees (encounters).}
  \label{fig:example_diags}
\end{figure}

We are now ready to present the mathematical definition of the
principal diagram.  Examples of unitary and orthogonal principal
diagrams are shown in Figs.~\ref{fig:encounter_untwist_prin} and
\ref{fig:example_diags}; the conditions entering the definitions are
discussed at length in the first part of the paper \cite{bk13a}.  When
comparing Figs.~\ref{fig:third_moment_example} and
\ref{fig:example_diags} note the shortened leaf labels.

\begin{definition}
  \label{def:unit_diagram}
  The \emph{unitary principal diagram} is a unicellular orientable map
  satisfying the following:
  \begin{enumerate}
  \item There are $t$ vertices of degree 1 (\emph{leaves}) labelled
    with symbols $1,\ldots, t$ and $t$ leaves labelled
    with symbols $\wb{1},\ldots, \wb{t}$.
  \item All other vertices have even degree greater than $2$.
  \item \label{enum:boundary} A portion of the boundary running from
    one leaf to the next is called a \emph{boundary segment}.  Each
    leaf $j$ is incident to two boundary segments, one of which is a
    segment running to the leaf $\wb{j}$ and the other running to the
    leaf $\wb{j-1}$.  The segments are given direction $j \to
    \wb{j}$ and $j\to\wb{j-1}$ and marked by solid and dashed
    lines correspondingly.  The following conditions are satisfied:
    \begin{enumerate}
    \item each part of the boundary is marked exactly once,
    \item \label{enum:two_sides} each edge is marked solid on one side
      and dashed on the other, both running in the same direction.
    \end{enumerate}
  \end{enumerate}
\end{definition}

Here \emph{unicellular} means that the diagram has one face, i.e.\ its
boundary is connected.  We take the operation $j-1$ to be cyclic:
$1-1=t$.  The leaves labelled $1$, \ldots, $t$ and $\wb{1}$, \ldots,
$\wb{t}$ we still call $i$-leaves and $o$-leaves correspondingly.

The conditions that make a valid orthogonal diagram are almost
identical to the unitary case.  The only significant difference is
that trajectories $\gamma$ and $\gamma'$ do not have to run in the
same direction.

\begin{definition}
  \label{def:orth_diagram}
  The \emph{orthogonal principal diagram} is a
  locally orientable map satisfying the following:
  \begin{enumerate}
  \item There are $t$ leaves labelled with symbols $1,\ldots, t$ and
    $t$ leaves labelled with symbols $\wb{1},\ldots, \wb{t}$.
  \item All other vertices have even degree greater than $2$.
  \item Each leaf $j$ is incident to two boundary segments, one of
    which runs to the label $\wb{j}$ and is marked solid,
    and the other runs to $\wb{j-1}$ and is marked dashed.  Each edge
    is marked solid on one side and dashed on the other.
  \end{enumerate}
\end{definition}

If the two boundaries of an edge are marked as running in the same
direction, this edge is called \emph{unitary}, otherwise it is
\emph{orthogonal}.  A unitary diagram has only unitary edges, while an
orthogonal diagram can have either.  A vertex is called \emph{unitary} if
all edges emanating from it are unitary.  Note that if we perform a
boundary walk of the diagram, the sides of a unitary edge will be
traversed in opposite directions, while the orthogonal edge will be
traversed in the same direction.

Finally, we formalize the notion of ``untying'' (it is explored in
more detail in the Appendix).

\begin{definition}
  \label{def:untiable}
  A vertex of even degree is called \emph{untieable} (i.e. ``can be
  untied'') if every second edge emanating from it leads directly to
  a leaf.  If these leaves all have $i$-labels, the vertex is called
  $i$-untieable.  If these leaves all have $o$-labels, the vertex is called
  $o$-untieable.
\end{definition}

For example, the lower right vertex in Fig.~\ref{fig:example_diags}(a)
is $i$-untieable, while the upper vertex in
Fig.~\ref{fig:example_diags}(b) is $o$-untieable.  It can also happen that
every second edge leads to leaves with a mix of $i$- and $o$-labels,
but only if the diagram is orthogonal (the lower vertex in
Fig.~\ref{fig:example_diags}(b) is an example). 

\begin{figure}[t]
 \includegraphics[scale=0.8]{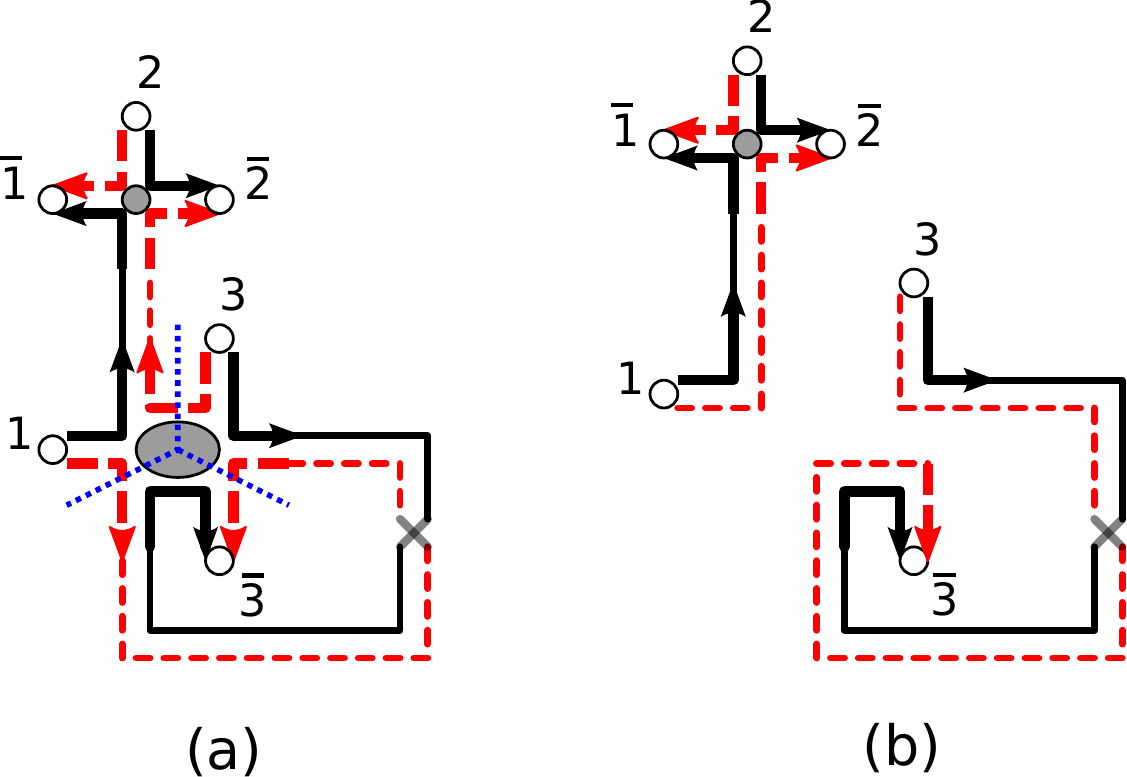}
 \caption{Untying a vertex of degree 3.}
 \label{fig:untying3_TR_short}
\end{figure}

Having defined what an untieable vertex is, we explain the operation
of untying, using the example of Fig.~\ref{fig:untying3_TR_short}.  An
untieable vertex of degree $2m$ is untied by cutting it into $m$
parts, preserving the solid boundary segments.  The dashed boundary
segments are then reconnected as necessary.  The semiclassical meaning
of an untieable vertex is an encounter that \emph{can} happen in a
lead: because of the last rule in Definition~\ref{def:Essen_ansatz},
the contribution of a diagram with an encounter (vertex) happening in
the lead is equal to the contribution of this diagram with the said
vertex untied.

Our summation over semiclassical diagrams will be organized by
grouping the contributions of a principal diagram together with its
untied versions.

\subsection{Contribution of a unitary diagram}
\label{sec:contrib_unit}

By following the rules in Definition~\ref{def:Essen_ansatz}, the
contribution of a principal diagram to the $n$-th moment $M_n(X)$ is
given by $(-1)^{\V} \Ni^n \No^n / N^{\E-\V}$, where $\E$ is the number
of edges of the diagram and $\V$ the number of internal vertices.
Again $\Ni\times\No$ is the size of $X^{\T}$.  Having defined the
untyings, we can now consider the contributions of the untied diagrams
and for this we first return to Example~\ref{exm:organization1}.

\begin{figure}[t]
  \includegraphics[scale=0.5]{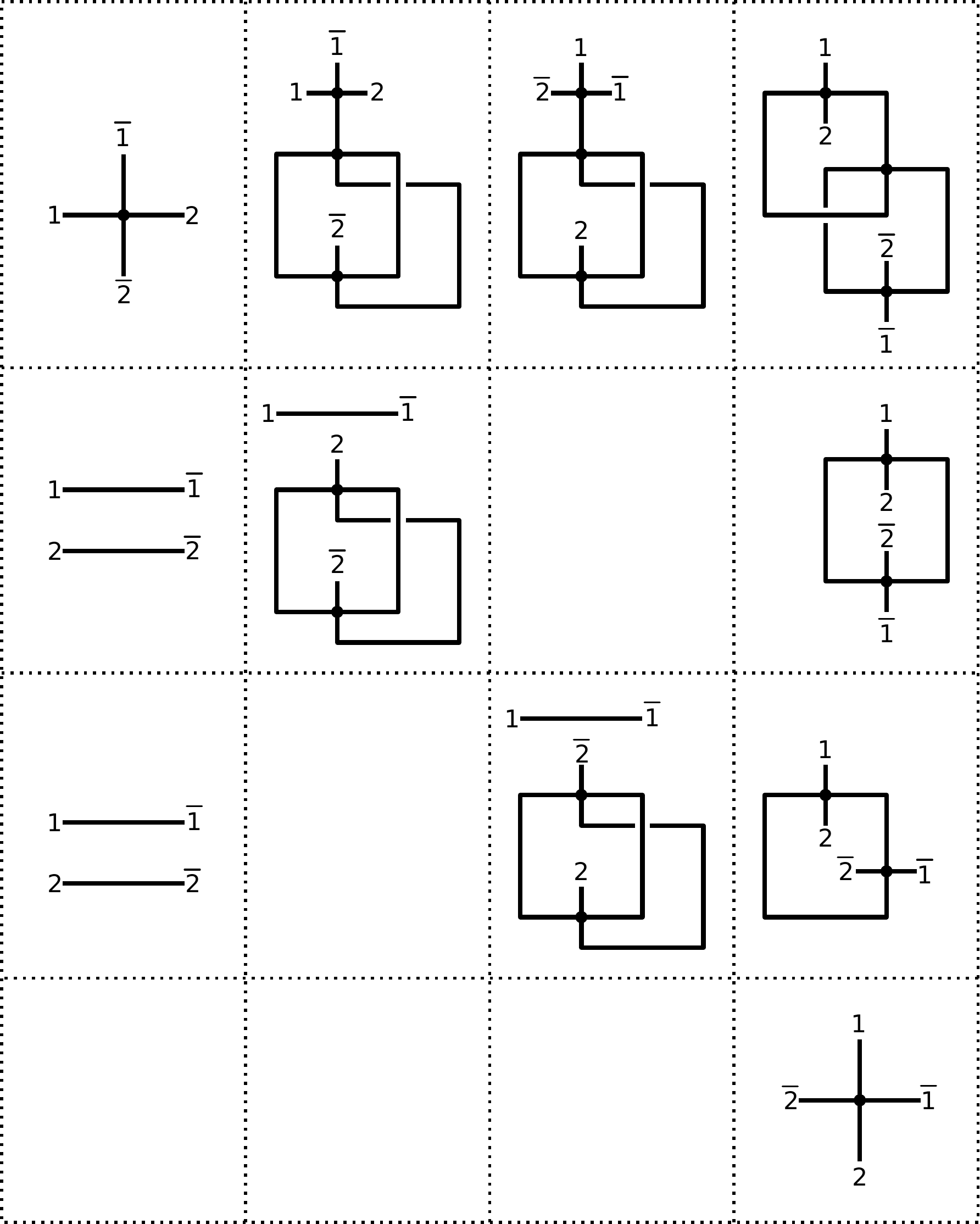}
  \caption{Some diagrams contributing to the correlator in Example~\ref{exm:organization2}.  
  The top row contains principal diagrams while untying their nodes leads to the diagrams below; see 
  main text.}
  \label{fig:principal}
\end{figure}

\begin{example}
  \label{exm:organization2}
  Similarly to equation~(\ref{eq:V_with_delta}), we reorganize the
  \emph{semiclassical} contributions to the moment $M_2(X)$ as
  \begin{equation}
    \label{eq:M2}
    M_2(X) = \Ni^2 \No^2 D^{\U}(\tau) + \Ni \No^2 D^{\U}( (1\, 2)\tau)
    + \Ni^2 \No D^{\U}(\tau(1\, 2)) + \Ni \No D^{\U}((1\, 2)\tau (1\, 2)),
  \end{equation}
  where $D^U(\tau)$ is the contribution of all unitary principal
  diagrams, $D^U((1\, 2)\tau)$ is the contribution of the principal
  diagrams after untying a vertex with leaf labels $i_1$ and $i_2$ and
  so on.  In Figure~\ref{fig:principal} we have several diagrams
  contributing to the sum.  They are arranged in the following manner.
  The four rows list diagrams contributing to the terms
  $D^{\U}(\tau)$, $D^{\U}((1\, 2)\tau)$, $D^{\U}(\tau(1\, 2))$ and
  $D^{\U}((1\, 2)\tau (1\, 2))$ correspondingly (top to bottom).  The
  diagrams in the lower three rows are the results of untying the
  diagram in the top row.  For example, the diagrams in the second row
  are the result of $i$-untying the diagram above it and are accounted
  for in the term $D^{\U}((1\,2)\tau)$.  Similarly, the diagrams in
  the third row are the result of $o$-untying the top diagram and
  contribute to $D^U(\tau(1\,2))$.  For the final row we $i$-untie one
  vertex and $o$-untie the other.  Some untyings are not possible and
  the corresponding positions are left empty.

  Note that we are essentially using $\tau$ as a placeholder symbol
  with the meaning ``principal diagram''.  If one chooses to delve
  deeper into the combinatorics of semiclassical diagrams (as done in
  \cite{bk13a} and the Appendix), $\tau$ takes the meaning of the
  \emph{target permutation}.  For unitary principal diagrams,
  $\tau=(1\ 2\ \ldots\ t)$ and the operation of untying corresponds to
  the actual multiplication of permutations.  This is explored in
  detail in the Appendix.

  We now observe that the contributions of all diagrams in a given
  column are of the same order (taking the prefactors in \eqref{eq:M2}
  into consideration).  For example the contributions of the last
  column are
  \begin{equation*}
    -\frac{\Ni^2 \No^2}{N^5} + \frac{\Ni \No^2}{N^4} + \frac{\Ni^2
      \No}{N^4} - \frac{\Ni \No}{N^3}.
  \end{equation*}
\end{example}

\begin{example}
\label{unitaryfigureexample}
  The diagram of Figure~\ref{fig:example_diags}(a) and its untied
  version give the contribution
  \begin{equation}
\label{unitaryfigureexampleeqn}
    -\frac{\Ni^3 \No^3}{N^7} + \frac{\Ni^2 \No^3}{N^6}.
  \end{equation}
  Note that we only untie the vertices that are ``untieable'' in the
  original diagram.  For example, the lower left vertex of this
  diagram \emph{becomes} untieable after untying the lower right
  vertex, but it is not a part of this particular sum.
\end{example}

To summarize, if a $2m$-vertex of a unitary diagram becomes untied,
its contribution is missing one vertex factor of $(-N)$, $m$ edge
factors of $1/N$ and there is only one factor of $N_j$ where before
there were $m$.  To include the contribution of the untied diagrams 
with the principal diagram, we multiply the contribution of the principal 
diagram by the factor
\begin{equation}
\left(1 - \frac{N^{m-1}}{N_j^{m-1}}\right) ,
\end{equation}
for each $2m$ vertex which can be untied, where $N_j$ depends on 
whether the vertex is $i$- or $o$-untied and is simply the number of channels 
in the corresponding lead.

Carrying on Example~\ref{unitaryfigureexample}, we then have
\begin{example}
  The diagram of Figure~\ref{fig:example_diags}(a) and its untied
  version together give the contribution
  \begin{equation*}
    -\frac{\Ni^3 \No^3}{N^7}\left(1 - \frac{N}{\Ni}\right) ,
  \end{equation*}
which is \eqref{unitaryfigureexampleeqn}.
\end{example}

We also mention that the contributions listed above are valid both for
the transmission moments, where $X$ is the off-diagonal matrix $\bs{t}$ in
\eqref{scatmatpartseqn} [with $\Ni=N_1$ and $\No=N_2$], and for the 
reflection moments (where $X$ is the diagonal matrix $\bs{r}$).  
In the latter case we additionally have $\Ni=\No=N_1$.

\subsection{Contribution of an orthogonal diagram}
\label{sec:contrib_orth}

The situation is somewhat different in the orthogonal case.  If the
vertex is purely $i$- or $o$-untieable, the contribution adjustment is exactly the same
as in the unitary case.  However, if the leaf labels involve a mixture of
labels of the two types, then the corresponding Kronecker delta [see
Equation~\eqref{eq:V_with_delta0}] mixes $i$ and $o$ indices.  For
transmission moments, where the incoming and outgoing channels are in 
separate leads, those cannot possibly coincide and the
corresponding untying produces 0 additional contribution.  When calculating
reflection moments, such ``mixed'' untieable vertices do contribute
and their contribution is calculated according to the rules above
(with the understanding that $\Ni=\No$).  Namely, the contribution of
the untied diagram is divided by $-\Ni^{m-1} / N^{m-1}$.

\begin{example}
  Consider the diagram of Fig.~\ref{fig:example_diags}(b).  The top
  vertex is $o$-untieable (if $o_1$ and $o_2$ are the same
  channel), while untying the lower vertex requires that
  $i_1$, $i_3$ and $o_3$ be the same channel and therefore in the same lead.  This is possible
  only in the input and output leads coincide, i.e.\ we are considering
  a reflection quantity.  The total contribution of this diagram,
  viewed as the principal diagram, is
  \begin{equation*}
    \frac{N_1^3 N_2^3}{N^6} \left(1 - \frac{N}{N_2} \right) , 
  \end{equation*}
  to the (third) transmission moment and 
  \begin{equation*}
    \frac{N_1^6}{N^6} \left(1 - \frac{N}{N_1} \right) \left(1 -
      \frac{N^2}{N_1^2} \right) ,
  \end{equation*}
  to the reflection moment.
\end{example}

\section{From principal diagrams to base structures}
\label{sec:basestructures}

Having understood how to evaluate the contribution of a particular
principal diagram and its untied version, we now turn to the question of generating the
diagrams.  Our aim eventually is to evaluate $M_n$ for any $n$, but only
to several leading orders of $1/N$, assuming $N_1\sim N_2\sim N$.

As mentioned in Section~\ref{sec:contrib_unit}, the contribution of a
principal diagram to the $n$-th moment is $(-1)^{\V} \Ni^n \No^n /
N^{\E-\V}$, where $\E$ and $\V$ are the number of edges and internal 
vertices of the diagram respectively.  Denoting the total
number of vertices (including the leaves) by $\Vt$ and noting that $\Vt = \V + 2n$, we see
that the order of the contribution is $1/N$ to the power $\E-\Vt$.  The
untyings of the principal diagram contribute at the same order.

Since the target permutation of the principal diagram is (see
Remark~\ref{remark:unitsubsetorth} in the Appendix) the palindromic
grand cycle $\tau=(1\,2\ldots n)(\wb{n} \ldots \wb{2} \, \wb{1})$, the
boundary is connected and the diagrams are \emph{unicellular}
(i.e.~have one face).  The \emph{genus} of an orientable map is
defined as the smallest genus of a surface on which the map can be
drawn without self-intersection.  Recalling that the genus $g$ of
unicellular orientable maps can be found as
\begin{equation}
  \label{eq:genus}
  2g = 1  + \E - \Vt ,
\end{equation}
the order of a diagram's contribution is $1/N$ to the power $2g-1$.
An asymptotic expansion in $1/N$ is then a type of genus expansion,
familiar from Gaussian ensembles and their applications
\cite{Zvo_mcm97}.  The genus of an orientable map must be integer,
however, if we take equation~(\ref{eq:genus}) as the definition in the
non-orientable case, orthogonal maps can have half-integer ``genus''
(there is a notion of \emph{demigenus} for non-orientable surfaces,
which is an integer and coincides with our value $2g$).

Our task is complicated by the fact that we would like to obtain
moments of arbitrary order $n$.  Thus our typical diagram has a low
genus and many vertices.  This suggests that we can
enumerate the eligible diagrams by planting trees (which provide many
vertices at no cost to genus) onto base structures that have the
required genus.

\begin{definition}
  A \emph{base structure} is a unicellular map with no vertices of
  degree 1 or 2 and with a labelled ``starting'' edge-side and
  specified direction.
\end{definition}

It is easy to see that the number of possible base structures
contributing at a given order is finite.  Indeed, since the minimal
vertex degree is 3, the number of edges can be estimated as $\E > 3\Vt/2$
and therefore, from \eqref{eq:genus}, the number of vertices
is bounded by $2(2g-1)$.

\begin{remark}
  Another name for base structures in the literature is ``schemes'',
  see \cite{ChaMarSch_siamjdm09,Cha_ptrf10}.
\end{remark}

In the sections that follow we describe the algebraic procedures for
generating the base structures and planting trees.  Before we do so,
we present several examples that illustrate the main ideas which we develop 
further in Section~\ref{sec:algoapproach}.

\begin{figure}[t]
  \includegraphics[scale=0.7]{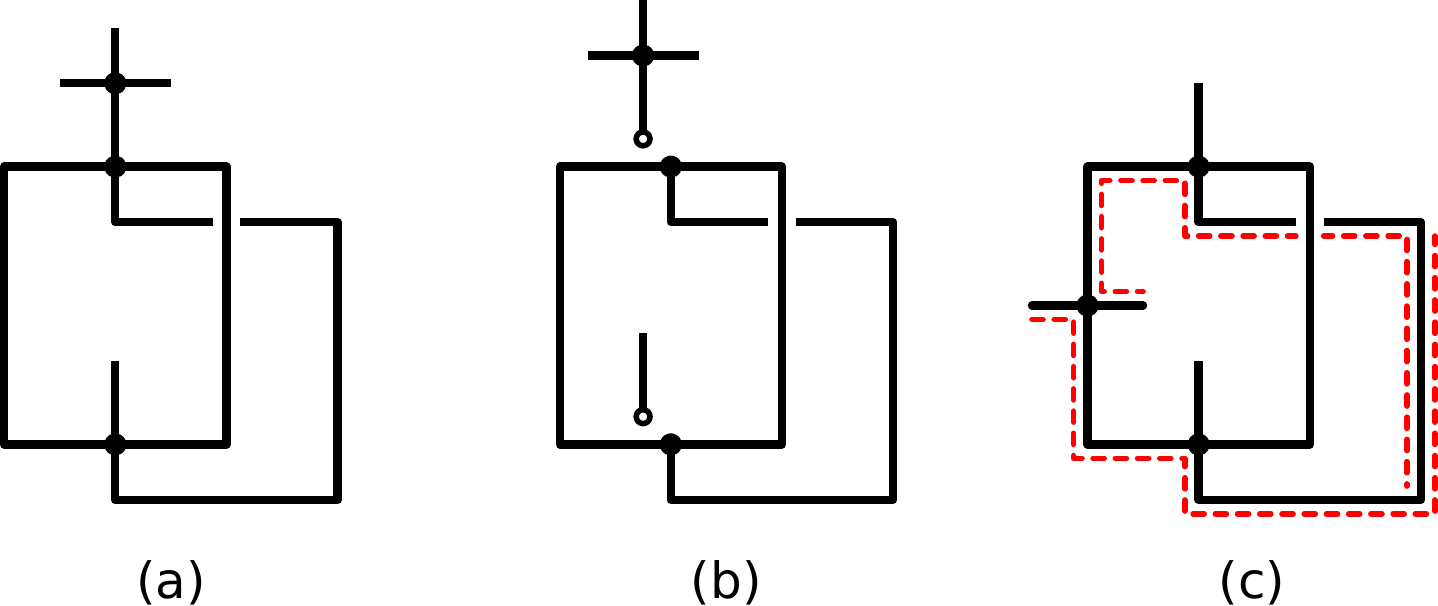}
  \caption{(a) An example of a diagram appearing in
    Fig.~\ref{fig:principal} and (b) its decomposition into a base
    structure and (rooted) trees.  (c) An example of invalid diagram
    and an attempt to label its boundary segments (only dashed boundary
    segments are shown): some edges will be labelled the same on both
    sides violating one of the requirements of
    Definition~\ref{def:unit_diagram}.}
  \label{fig:grafting_ex1}
\end{figure}

Fig.~\ref{fig:grafting_ex1}(a)-(b) shows an example of a diagram, its
base structures and the trees.  Reversing the process, we will plant
trees with internal vertices of even degrees greater than 2.
Obviously we have to plant enough trees to make all vertices on the
base structure have even degree.  However, as the example of
Fig.~\ref{fig:grafting_ex1}(c) shows, this is not sufficient to
generate a valid diagram.  The obstacle is the requirement that the
solid and dashed trajectories match along the boundary to satisfy
Definitions~\ref{def:unit_diagram} or \ref{def:orth_diagram}.  It is
possible to theoretically characterize a map whose boundaries can be
properly labelled solid and dashed as required.  Rather than doing
this, however, we will describe a construction method which generates
only valid diagrams.

\begin{figure}[t]
  \includegraphics[scale=0.7]{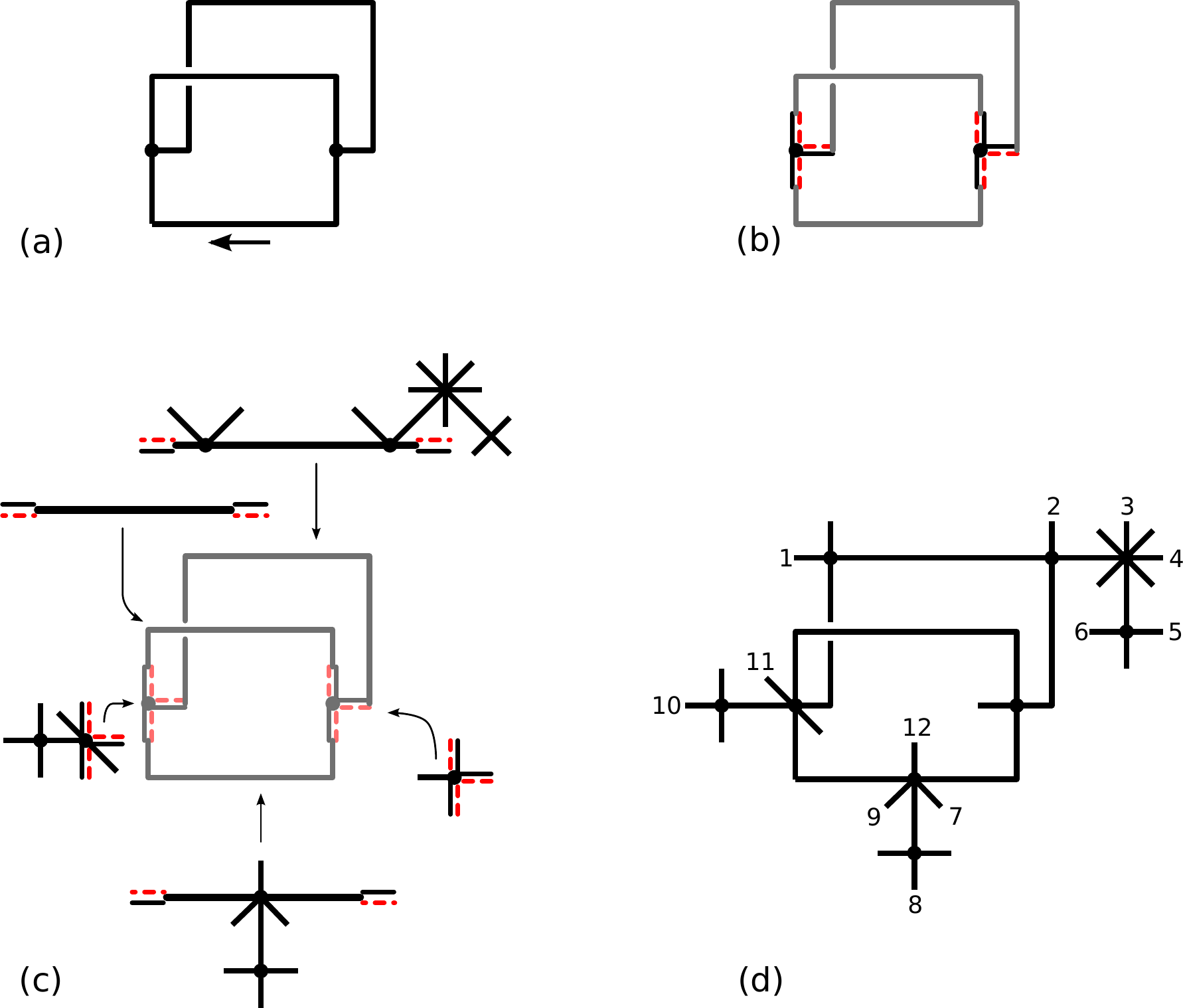}
  \caption{(a) The base structure, (b) pre-marking the edge-ends, (c)
    planting trees on vertices and edges, (d) labeling the leafs.  To
    avoid clutter only the leaf labels without bars are
    shown.}
  \label{fig:grafting}
\end{figure}
 
We first outline the method using the example of
Fig.~\ref{fig:grafting}.  Starting with a base structure (details in
section~\ref{sec:base}) we pre-label the stubs of edges around every
vertex with dashed and solid lines.  By a ``stub'' of an edge we
understand a small part of edge attached to the vertex.  The
pre-labeling can be done in arbitrary manner, provided the lines are
different on the two sides of each edge, see
Fig.~\ref{fig:grafting}(b).  Then we plant rooted trees
(section~\ref{sec:trees}) on edges and vertices.  The parity of the
number of trees and their type is fully determined by the pre-marking
(sections~\ref{sec:algoedges} and \ref{sec:algovertices}).  The
contribution of the pre-marked diagram to the total sum is expressible
as the product of the contributions of its constituent parts: edges
and vertices.  Finally, we will sum the contributions over all
possible pre-markings.

\subsection{Generating base structures}
\label{sec:base}

The semiclassical diagrams are drawn as ribbon graphs with the edges
fattened to have two sides.  We now present a combinatorial
description of the base structures which is a slightly modified version
of Tutte's axiomatization \cite{JacVis_Atlas,Tutte_GraphTheory}.

From the definition of the base structure, we obtain the canonical boundary walk
which starts at the marked edge in the marked direction and passes
every edge twice (once on each side).  As we go along the boundary, we
label the edge-sides with numbers $1, \ldots 2m$, where $m$ is the
number of edges in the base structure.  We also mark the direction
of each edge-side.  The reversal of an edge-side (i.e.\ the same
edge-side running in the opposite direction) is denoted by the same
symbol with a bar.  Therefore the reversal of the canonical boundary
walk passes the edge-sides $\wb{2m}, \ldots \wb{1}$.  It turns out
that a base structure with such labelling is uniquely specified by the
pairing (matching) of the labels on the opposite sides of the edges.

To generate base structures with $m$
edges, we consider permutations on the set 
\begin{equation*}
  Z_{2m} = \{1,\ldots, 2m, \wb{2m}, \ldots, \wb{1}\} , 
\end{equation*}
of $4m$ elements.  The permutation
\begin{equation}
  \label{eq:reversal}
  \Tt = (1\, \wb{1})(2\, \wb{2})\cdots (2m\, \wb{2m}) ,
\end{equation}
encodes the operation of reversal while the \emph{face permutation}
\begin{equation}
  \label{eq:face_perm}
  \phi = (1\, 2 \ldots 2m)(\wb{2m} \ldots \wb{2}\, \wb{1}) , 
\end{equation}
corresponds to the canonical boundary walk of the unique face of the
map and its reversal.  With these pieces of data fixed, the unicellular
map is described by one permutation.

\begin{definition}
  \label{def:orth_map_def}
  A unicellular map in canonical form is a permutation $\varepsilon$
  that 
  \begin{itemize}
  \item  is a fixed-point free involution (i.e.\ has only cycles of
    length 2),
  \item has no cycles of the form $(x\, \wb{x})$,
  \item commutes with $\Tt$: $\Tt \varepsilon
    = \varepsilon \Tt$.
  \end{itemize}
\end{definition}

The cycles of $\varepsilon$ correspond to the matching of different
sides of the edges.  For example, a cycle of $(1\ 3)$ means that one
edge has sides numbered $1$ and $3$ running \emph{in the opposite
  directions}.  Then, the reversals $\wb{1}$ and $\wb{3}$ must also be
matched.  This is ensured by the commutativity requirement: $\Tt(1\
3)\Tt^{-1} = (\wb{3}\, \wb{1})$.  

The cycle of the form $(1\, \wb{3})$ [and its counterpart $(3\,
\wb{1})$] would denote an edge with sides $1$ and $3$ running \emph{in
  the same direction}.  There are no such edges in an orientable map.
An edge of the form $(j\, k)(\wb{k}\, \wb{j})$ we will call a
\emph{unitary edge}, while the edge of the form $(j\, \wb{k})(k\,
\wb{j})$ will be referred to as an \emph{orthogonal edge}.  We stress
that a diagram contributing to an orthogonal (i.e.\ with TRS) quantity
may contain some unitary edges.  It may even contain only unitary
edges: a unitary diagram contributes in both cases.

The permutation
\begin{equation}
  \label{eq:orth_vertices}
  \nu = \phi \varepsilon ,
\end{equation}
is called the \emph{vertex permutation}.  Each vertex of the map corresponds
to two cycles that list the edge-sides leaving the vertex.  One cycle
has the edge-sides that keep their edge to their left, listed
anticlockwise around the vertex.  The other lists the edge-sides that
keep their edges to their right, in the clockwise order around the
vertex.  Naturally, the base diagrams are unicellular maps whose
vertex permutation only has cycles of length 3 or higher.

\begin{figure}[t]
  \centering
  \includegraphics[scale=0.7]{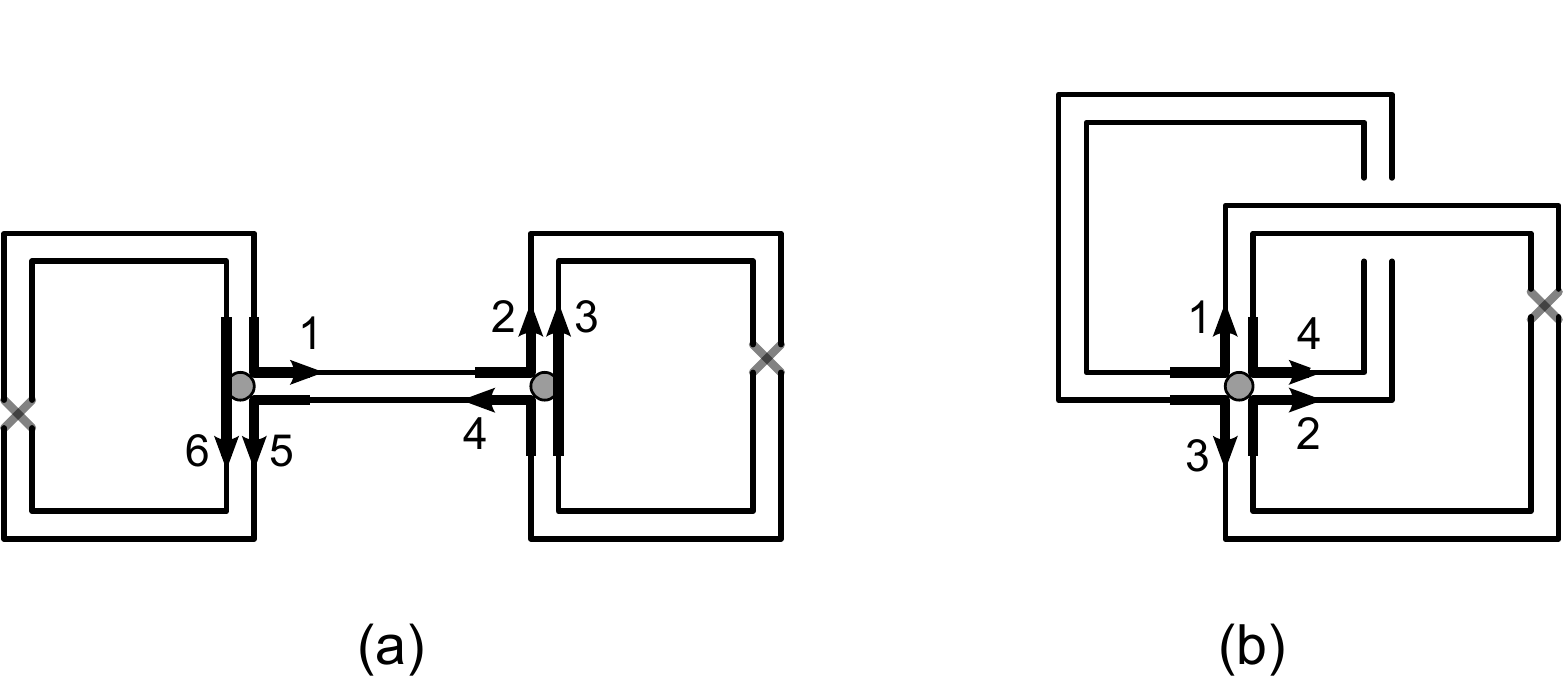}
  \caption{Two examples of orthogonal base diagrams.  The diagram in
    (a) has one unitary edge (in the middle) and two orthogonal
    edges.  The diagram in (b) has one orthogonal edge (left) and one
    unitary edge (right).  Note that the presence of a twist in the
    edge does not mean the edge is orthogonal.}
  \label{fig:base_diagO}
\end{figure}

\begin{figure}[t]
  \centering
  \includegraphics[scale=0.7]{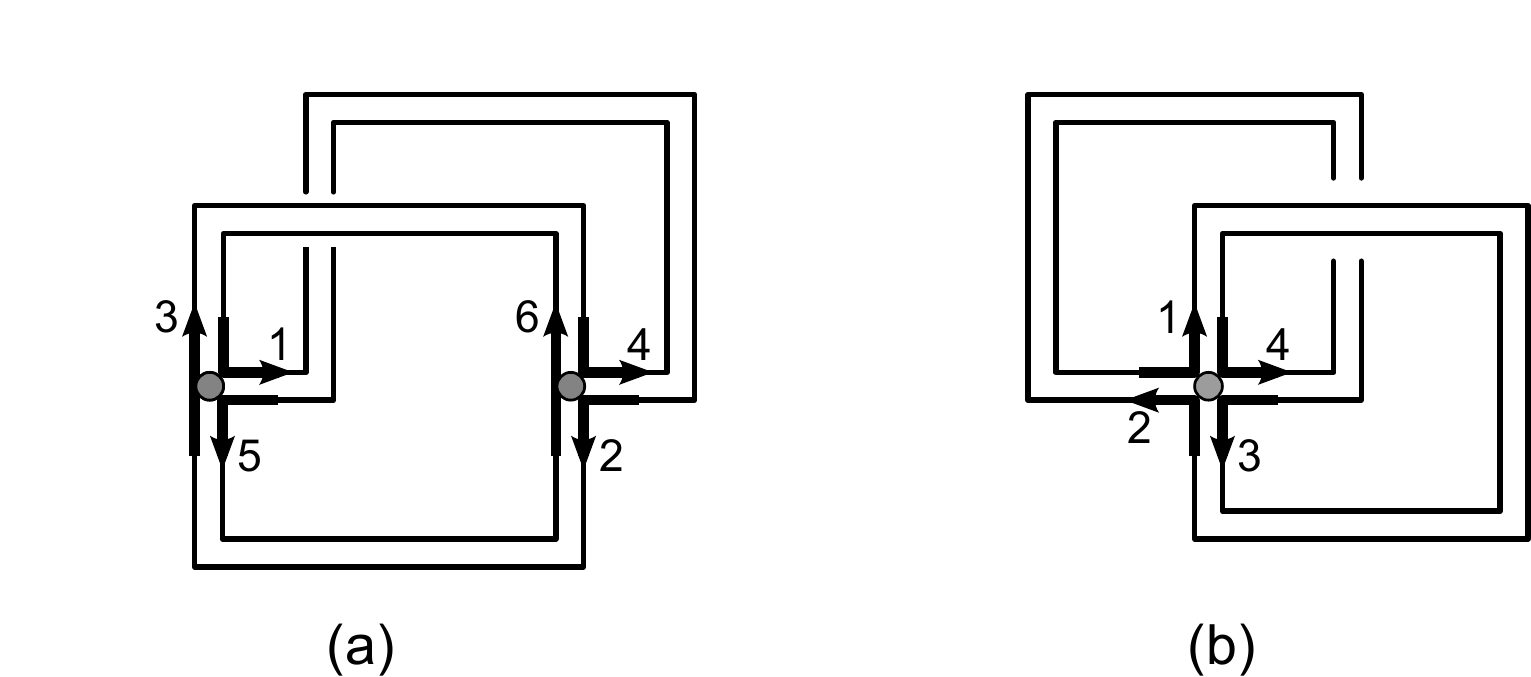}
  \caption{Two examples of unitary base diagrams.  These are the only
    unitary diagrams of genus $1$.} 
  \label{fig:base_diagU}
\end{figure}

\begin{example}
  \label{exm:base_diagO}
  The map from Fig.~\ref{fig:base_diagO}(a) can be represented as
  \begin{equation*}
    \varepsilon = (1\, 4)(2\, \wb{3})(5\,
    \wb{6}) (\wb{5}\, 6) (\wb{2}\,
    3) (\wb{1}\, \wb{4}), \quad \mbox{with} \quad
    \nu = (1\, 5\, \wb{5})(\wb{4}\, \wb{6}\, 6)(2\,
    \wb{2}\, 4)(\wb{1}\, \wb{3}\, 3),
  \end{equation*}
  while the map of Fig.~\ref{fig:base_diagO}(b) can be written as
  \begin{equation*}
    \varepsilon = (1\, 3)(2\, \wb{4}) (\wb{2}\,
    4)(\wb{1}\, \wb{3}), 
    \quad \mbox{with} \quad \nu=(1\, 4\, \wb{1}\, \wb{2})(2\, \wb{3}\,
    \wb{4}\, 3). 
  \end{equation*}
\end{example}

When performing a computation, we choose a canonical way to order the
cycles in the permutation $\varepsilon$.  For example, we order the
elements of $Z_{2m}$ by mapping $\wb{x}$ to $x+2m$ for all $x\in
\{1,\ldots,2m\}$ and order the cycles in a palindromic fashion
\begin{equation*}
  \varepsilon = (s_1\,r_1) \, (s_2\,r_2) \cdots (s_\V\,r_\V)
  (\wb{s_\V}\,\wb{r_\V}) \cdots (\wb{s_2}\,\wb{r_2}) \,
  (\wb{s_1}\,\wb{r_1}), 
\end{equation*}
with the ordering conditions
\begin{equation*}
  s_j < r_j,\quad s_j < \wb{s_j}, \quad s_j < \wb{r_j}, \quad s_j \leq s_{j+1}.  
\end{equation*}
With the additional requirement $r_j \neq \wb{s_j}$ the above
palindromic permutations automatically satisfy all the conditions of
Definition~\ref{def:orth_map_def}.  Calculating the permutation $\nu$
we establish how the edges are connected to the vertices.  At this
point we exclude the diagrams that have cycles of length 1 or 2 in the
permutation $\nu$.  Next we calculate the semiclassical contribution
of the base structure following the prescriptions explained in Section~\ref{sec:algoapproach}.

With broken TRS, all the edges must be traversed on both sides by semiclassical 
trajectories travelling in the same direction.  Or, equivalently,

\begin{definition}
  A \emph{unitary} base structure is an orientable base structure. 
\end{definition}

The cycles of $\varepsilon$ can then only involve pairs of labels either 
both with bars or both without bars.  Removing the redundant half of 
$\varepsilon$ involving bars, we return to the standard 
definition:

\begin{definition}
  An orientable map of size $m$ is a triple $(\wt{\varepsilon},\wt{\nu},\wt{\phi})$ of
  permutations of size $2m$ such that all cycles of $\wt{\varepsilon}$ have
  length 2 and $\wt{\nu}\wt{\varepsilon} = \wt{\phi}$.
\end{definition}

For the unitary base structures we have $\wt{\phi} = (1\, 2\ldots
2m)$ and we again exclude diagrams with vertices of degree 1 and 2.

\begin{example}
  \label{exm:base_diagU}
  The maps from Fig.~\ref{fig:base_diagU} can be represented as
  \begin{equation*}
    \wt{\varepsilon} = (1\, 4)(2\, 5)(3\,
    6), \quad \mbox{with} \quad
    \wt\nu = (1\, 5\, 3)(2\, 6\, 4),
  \end{equation*}
  and
  \begin{equation*}
    \wt\varepsilon = (1\, 3)(2\, 4), 
    \quad \mbox{with} \quad \wt\nu=(1\, 4\, 3\, 2), 
  \end{equation*}
  where the vertices can be read off clockwise in Fig.~\ref{fig:base_diagU}.
\end{example}

As the size of the permutations is halved, the search for unitary base
structure is computationally more efficient than the search for the
orthogonal ones.  This allows us to go to a higher genus
(semiclassical correction order) in the case of broken TRS.  However,
if, for a given genus, the orthogonal base structures have already
been found, the unitary structures can be efficiently selected as a
subset of those.  In Table~\ref{tab:genus1_list} we list all
orthogonal base structures of genus $1$; the unitary base structures
are those whose permutation contains no bars, which were sketched in
Fig.~\ref{fig:base_diagU}.

\renewcommand{\arraystretch}{1.3}
\begin{table}[t]
  \centering
  \begin{tabular}{|c|c||c|c|}
    \hline
    $m$ & $\varepsilon$ & $m$ & $\varepsilon$ \\
    \hline
    $2$ & $(1\,3)(2\,4)$ & $3$ & $(1\,4)(2\,5)(3\,6)$ \\
    & $(1\, \wb{2})(3, \wb{4})$ && $(1\,\wb{2})(3\,6)(4\,\wb{5})$ \\
    & $(1\, 3)(2, \wb{4})$ && $(1\,\wb{3})(2\,5)(4\,\wb{6})$ \\
    & $(1\, \wb{3})(2, 4)$ && $(1\,4)(2\,\wb{3})(5\,\wb{6})$ \\
    & $(1\, \wb{4})(2, \wb{3})$ && $(1\,4)(2\,\wb{6})(3\,\wb{5})$ \\
    &&& $(1\,\wb{5})(2\,\wb{4})(3\,6)$ \\
    &&& $(1\,\wb{6})(2\,5)(3\,\wb{4})$ \\
    \hline
  \end{tabular}
  \vspace{3pt}
  \caption{Base structures of genus $g=1$ with $m$ edges.  Only half of the
    palindromic representation of $\varepsilon$ is given.}
  \label{tab:genus1_list}
\end{table}

To illustrate the difficulty of summation over the base structures, in
Table~\ref{tab:genus_number} we list the number of the base structures
of given genus $g$ and number of edges $m$.  In the unitary
(orientable) case these numbers have been studied, in particular, in
\cite{HarZag_im86}.  In the orthogonal (locally orientable) case, related
quantities have been considered in \cite{Ber_aam12}.

\begin{table}[t]
  \centering
  \begin{tabular}{|c|c|c|c||c|c|c|c|}
    \hline
    $g$ & $m$ & orth. & unit. & $g$ & $m$ & orth. & unit. \\ 
    \hline
    1 & 2 & 5 & 1 & 2 & 4 & 509 & 21 \\
    & 3 & 7 & 1 & & 5 & 4508 & 168 \\
    3/2 & 3 & 41 & & & 6 & 14235 & 483 \\
    & 4 & 198 & & & 7 & 20867 & 651 \\
    & 5 & 285 & & & 8 & 14516 & 420 \\
    & 6 & 128 & & & 9 & 3885 & 105\\
    \hline
  \end{tabular}
  \caption{Number of base structures at a given genus $g$ with $m$ edges.}
  \label{tab:genus_number}
\end{table}

\section{Summation over principal diagrams}
\label{sec:algoapproach}

Given a base structure we will now graft trees onto its edges and
vertices to create the principal diagrams.  

\subsection{Trees}
\label{sec:trees}

\begin{figure}[t]
  \centering
  \includegraphics{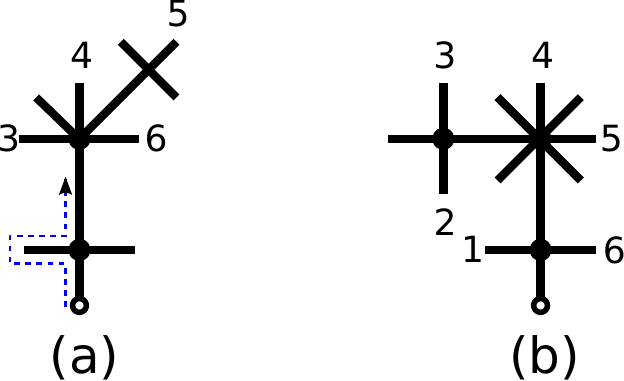}
  \caption{Examples of trees: (a) an $o$-tree and (b) an $i$-tree.
    The root is marked by the empty circle.  Only the $i$-leaves are
    labelled to avoid clutter.  A beginning of the boundary walk is
    shown by the dotted line.  In example (a), the leaf number 1 is
    located on some other part of the diagram, prior to the place
    where the tree is rooted.}
  \label{fig:trees}
\end{figure}

The leaves of grafted trees correspond to the incoming channels (with
labels from the set $\{1,2,\ldots,n\}$) and outgoing channels (with
labels from $\{\wb{1},\wb{2},\ldots,\wb{n}\}$).  For simplicity we
will refer to the incoming channel leaves as $i$-leaves and the
outgoing leaves as $o$-leaves.  The boundary walk of the trees
alternatively visits $i$ and $o$-leaves. There is an even number of
leaves altogether, but the \emph{root leaf} (which is where the tree
is to be attached to the base structure) is not labeled.  Thus an odd
number of leaves is labelled.  The trees with more $o$-leaves than
$i$-leaves will be called $o$-trees; their semiclassical contribution
will be denoted by $f$.  The contribution of the trees with more
$i$-leaves (``$i$-trees'') will be denoted $\fh$.  The exact form of
the contribution depends on the particular transport quantity that is
being considered and will be derived in
sections~\ref{sec:trees_transmission} and
\ref{sec:calculation_reflection}.
 
We mention that such rooted trees have also been used to find the leading
order moment generating functions for the transmission eigenvalues
\cite{bhn08} and the Wigner delay times \cite{bk10}.

\subsection{Edges} 
\label{sec:algoedges}

We now derive the contribution of an edge of a base structure on which
some trees have been grafted.  When trees are grafted at a point on
the edge, the point becomes a vertex.  To form a vertex of even
degree an even number of trees must be grafted.  The trees can be
placed on either side of the edge which creates two types of vertices:
\emph{odd vertices} with an odd number of trees attached to either
side (for example, the vertex on the lower edge of
Fig.~\ref{fig:grafting}), and \emph{even vertices} with an even
number of trees on either side (both vertices on the upper edge of
Fig.~\ref{fig:grafting}).

The semiclassical contribution of a vertex depends on $f$ and $\fh$ as
well as the exact transport quantity considered.  For now, we denote
by $A$ the contribution of an even node.  The odd nodes come in three
further subvarieties: those with a majority of $o$-trees attached,
those with a majority of $i$-trees and those with an equal number.
This last possibility occurs if and only if the edge is orthogonal
(i.e.\ traversed in the same direction by the boundary walk of the base
structure).  Their contributions will be denoted by $B_{o}$, $B_{i}$
and just $B$ correspondingly.

\begin{figure}[t]
  \centering
  \includegraphics[scale=0.5]{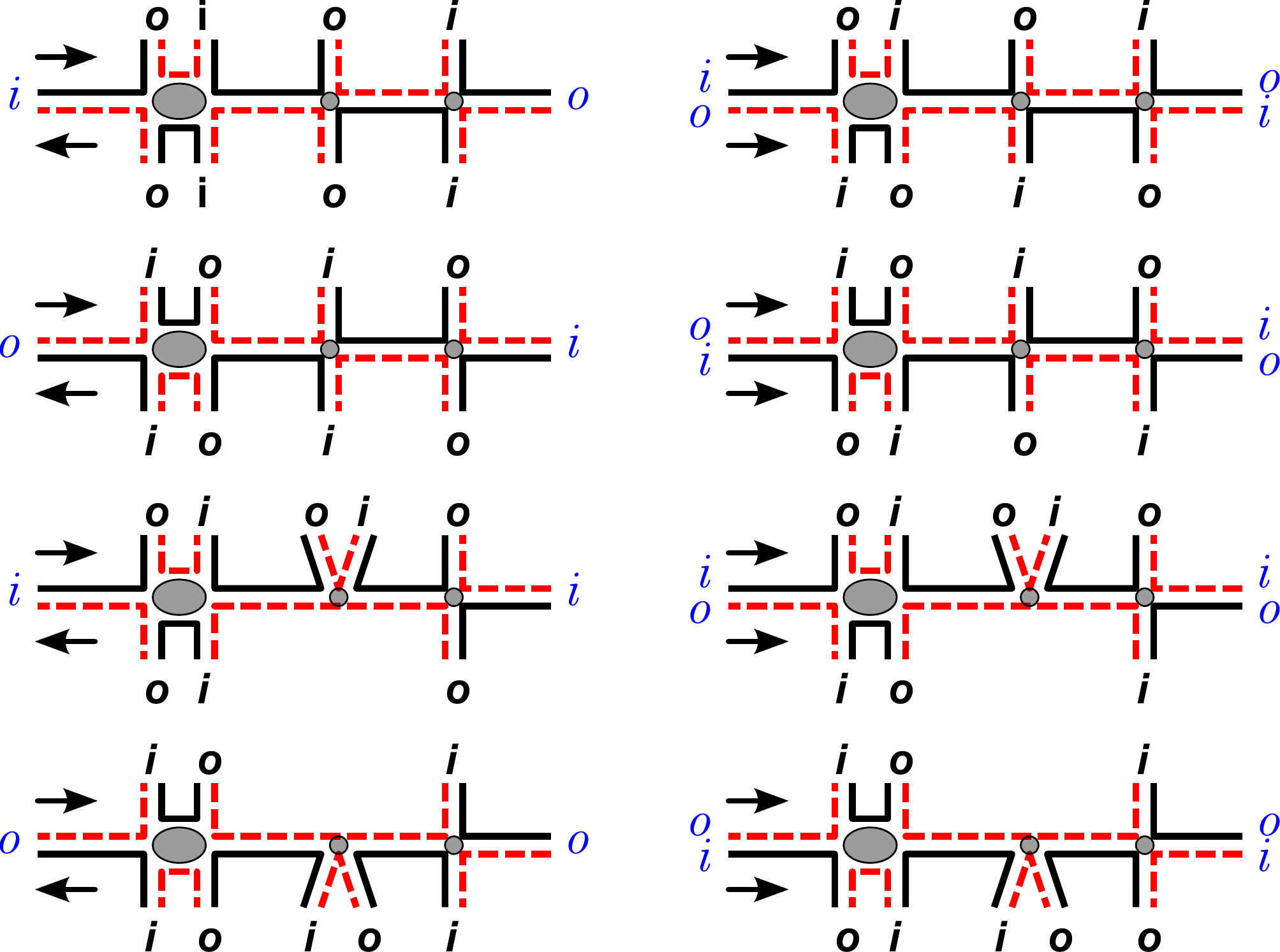}
  \caption{Examples of all possible types of pre-labelled edges.  The
    unitary edges are shown on the left, orthogonal are shown on the
    right (the direction of the boundary walk is indicated by arrows).}
  \label{fig:even_and_odd}
\end{figure}

After pre-marking of the edge ends with dashed and solid lines, 8
types of edge arise.  These depend on the pre-marking of the ends
(two types for each end) and on whether the edge is unitary or
orthogonal.  Examples of these types are given in
Fig.~\ref{fig:even_and_odd}.

\begin{figure}[t]
  \centering
  \includegraphics[scale=0.7]{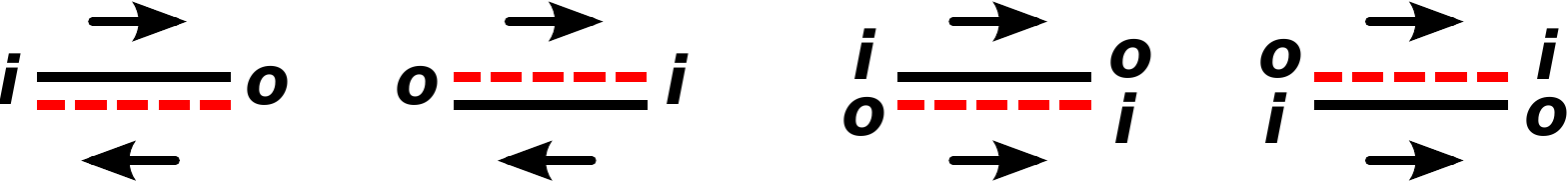}
  \caption{Rules for labeling edge ends: $i$ at the start of a solid
    segment or the end of a dashed segment; $o$ at the end
    of a solid or the start of a dashed segment.}
  \label{fig:edge_germs_io}
\end{figure}

We distinguish the different types using the labels that \emph{would}
be assigned to the edge ends.  This label depends on the direction of
the boundary walk along the edge: a boundary segment starts at $i$ and
ends at $o$, see Fig.~\ref{fig:edge_germs_io}.  It is important to
note that a solid segment runs \emph{along} the boundary walk, while
the dashed one runs in the \emph{opposite} direction.  Implementing
the above rule results in having one label per end for a unitary edge
but two labels per edge end for an orthogonal edge: one for each side.

Assigning the labels to the edge ends also preserves the alternation
of the $o$ and $i$ trees around the edge structure.  The edges on the
left side of Fig.~\ref{fig:even_and_odd} give contributions $\Eu(i,
o)$, $\Eu(o, i)$, $\Eu(i,i)$, $\Eu(o, o)$ listed top to bottom.  The
contributions $\Eu(i, o)$ and $\Eu(o, i)$ are equal, since their
configurations are related by the rotation by $\pi$.

The contributions of orthogonal edges is denoted by reading the
edge-end labels in the clockwise direction around the edge:
$\Eo(oi,oi)$, $\Eo(io,io)$, $\Eo(oi,io)$ and $\Eo(io,oi)$ for the
edges on the right side of Fig.~\ref{fig:even_and_odd} listed top to
bottom.  There are only two distinct contributions: two pairs are
related by top-bottom reflection, resulting in $\Eo(oi,oi) =
\Eo(io,io)$ and $\Eo(oi,io) = \Eo(io,oi)$.

\begin{figure}[t]
  \centering
  \includegraphics[scale=0.85]{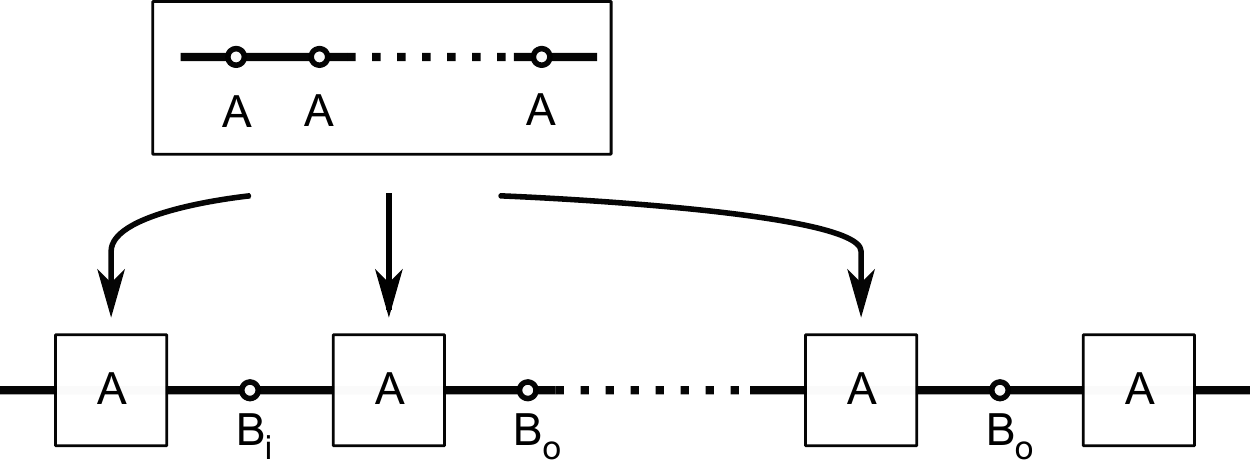}
  \caption{The structure of the edge $\Eu(o,i)$: odd nodes separated
    by sequences of even nodes.}
  \label{fig:edge_composition}
\end{figure}

We will now derive the contributions of a unitary edge in terms of the
already defined
quantities.  The structure of every edge is a sequence of
alternating odd nodes $B_{o}$ and $B_{i}$, separated by blocks of
even nodes, see Fig.~\ref{fig:edge_composition}.  Each block can have
any number of even nodes (or none at all), giving the contribution
\begin{equation*}
  y + y^2A + y^3 A^2 + \ldots  = \frac{y}{1-yA},
\end{equation*}
where $y$ is the semiclassical contribution of an edge in the diagram
(not to be confused with the ``composite'' edge of the base structure).  
From Definition~\ref{def:Essen_ansatz}, $y=1/N$ for the quantities we consider 
in section~\ref{sec:calculation}, though it differs in other physical situations.

The edge types $\Eu(o, i)$ and $\Eu(i, o)$ contain an equal number
of odd vertices $B_o$ and $B_{i}$, leading to
\begin{equation}
  \Eu(o, i) = \Eu(i, o) 
  = \sum_{n=0}^{\infty} \frac{B_{o}^nB_{i}^ny^{2n+1}}{(1-yA)^{2n+1}} 
  = \frac{y(1-yA)}{(1-yA)^2-y^2 B_o B_{i}}
\end{equation}
The $\Eu(o,o)$ edge has an extra odd $B_{i}$ vertex (and an extra
string of even nodes) and we have 
\begin{equation}
  \Eu(o,o) = \frac{yB_{i}}{(1-yA)} \Eu(o,i) 
  = \frac{y^2 B_{i}}{(1-yA)^2-y^2 B_o B_{i}}.
\end{equation}
Similarly, 
\begin{equation}
  \Eu(i,i) = \frac{y^2B_{o}}{(1-yA)^2-y^2 B_o B_{i}}.
\end{equation}

For the orthogonal edges, the difference with respect to unitary edges
is that the odd nodes are all of the same type with contribution $B$.
The edge $\Eo(oi,oi)$ has an even number of $B$ vertices, while
$\Eo(oi,io)$ has an odd number, leading to
\begin{equation}
  \Eo(oi,oi) = \Eo(io,io) = \frac{y(1-yA)}{(1-yA)^2-y^2B^2}, 
  \qquad \Eo(oi,io) = \Eo(io,oi) = \frac{y^2B}{(1-yA)^2-y^2B^2}.
\end{equation}
We remark that for the transport quantities we consider it turns out
that $B_iB_o = B^2$ which greatly simplifies the calculations.

\subsection{Vertices}
\label{sec:algovertices}

\begin{figure}[t]
  \centering
  \includegraphics[scale=0.85]{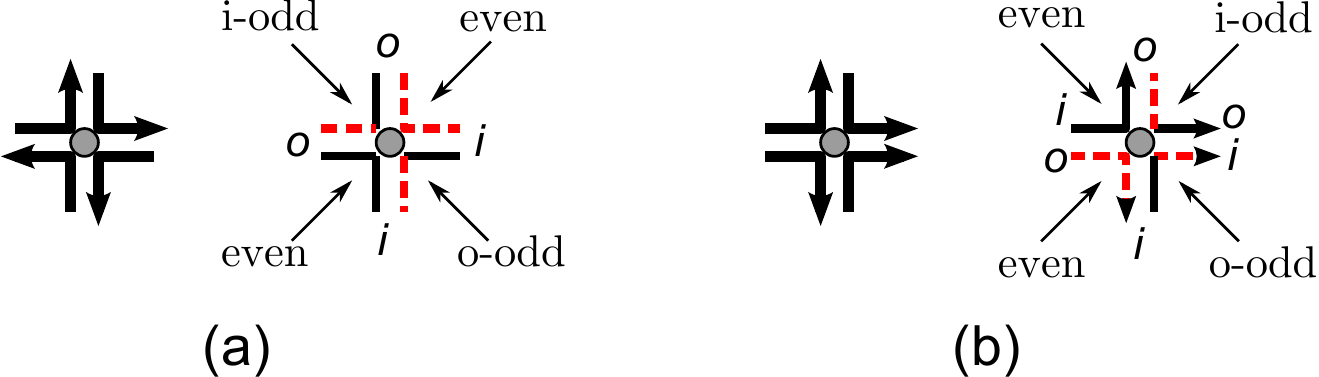}
  \caption{Examples of pre-labelled vertices, (a) unitary and (b)
    orthogonal.  The types of resulting sectors are indicated. }
  \label{fig:vertex_composition}
\end{figure}

Finally we can also graft trees onto the vertices of the base diagram.  
After the edge stubs of the base diagram have been pre-labelled, we
can assign labels to the edge stubs adjacent to a given vertex
according to the rules summarized in Fig.~\ref{fig:edge_germs_io}.
Knowing the labels we determine what type of trees can be planted in
the sectors between the existing edges.  There are three
possibilities: between labels $i$ and $i$ one has to plant an odd
number of trees, majority of them of type $o$; between labels $o$ and $o$
one plants an odd number of trees, majority of them type $i$; between
labels $i$ and $o$ one plants an equal number of $i$ and $o$ trees.
The resulting sectors will be referred to as $o$-odd, $i$-odd and even
correspondingly, see Fig.~\ref{fig:vertex_composition}.

The contribution of an even sector is thus
\begin{equation*}
  1 + f\fh + \left(f\fh\right)^2 + \ldots = \frac{1}{1-f\fh},
\end{equation*}
while the $o$ and $i$-odd sectors contribute
\begin{equation*}
  \frac{f}{1-f\fh} \qquad\mbox{and}\qquad
  \frac{\fh}{1-f\fh},
\end{equation*}
correspondingly.

Recording the labels of the edge stubs clockwise around a base diagram
vertex of degree $k$, we obtain the sequence $(b_1,\ldots,b_k)$.  The
semiclassical contribution of that vertex $V_k(b_1,\ldots,b_k)$ then
depends on the number of times $o$ follows $o$ (denoted by $p$) and
the number of time $i$ follows $i$ (denoted by $q$) in that sequence
(considered cyclically).  For example the code for the vertex in
Fig.~\ref{fig:vertex_composition}(b) is $oi,i,oi,o$ with $p=1$ and
$q=1$.

Finally, a special correction factor may arise due to the vertex
becoming untieable (see Definition~\ref{def:untieable}).  Since only
the planted trees can lead directly to a leaf, it is clear that the
vertex can only become untieable if all sectors are odd.  In addition
(when calculating the transmission moments), all sectors must have the
same type.  However, it is easy to see that the sectors on the two
sides of an orthogonal edge always have different type (if both odd).
Therefore, in the calculation of the transmission moments, the untying
of orthogonal vertices does not contribute (see
section~\ref{sec:contrib_orth}).  In the calculation of reflection
moments the type restriction becomes irrelevant and a vertex should
receive a correction factor whenever all sectors are odd.

To summarize, denoting the semiclassical contribution of the vertex (of the final
diagram) by $x$ and the untieable factor by $\chi$ (to be calculated
later), the contribution of a vertex is 
\begin{equation*}
  V_k(b_1,\ldots,b_k) = x \frac{f^q \fh^p}{(1-f\fh)^k} \chi.
\end{equation*}

\subsection{The algorithm} \label{sec:algoapproachalgo}

The contributions of edges and vertices are multiplicative: for a
given labelling of the edge stubs, we determine the contribution of
each constituent part of the base structure and multiply them together
to obtain the contribution of the pre-labelled base structure itself.
To obtain the total contribution of the base structure we sum over all
possible pre-labellings.  

In practical implementation, it is more convenient to assign the
symbols $i$ or $o$ to the ends of a unitary edge and symbols $io$ or
$oi$ to the ends of an orthogonal edge and then assign the opposite
values to the corresponding stubs of the vertices.  The unitary
diagrams are a subclass of the orthogonal ones, so we will concentrate
on the orthogonal case.

We will now describe the formal algorithm.  For a diagram with $m$
edges, introduce $4m$ variables $b_1, \ldots, b_{2m}, b_{\wb{1}},
\ldots, b_{\wb{2m}}$.  These  will take values in the set $\{i, o, io,
oi\}$.  We introduce two operations on this set, given by
\begin{align}
  \wt{i} &= o, & \wt{o} &= i, & \wt{io} &= oi, & \wt{oi} &= io, \\
  \wh{i} &= i, & \wh{o} &= o, &\wh{io} &= oi, & \wh{oi} &= io. 
\end{align}

We remind the reader that the base diagram is encoded by the
permutation $\varepsilon$ (which describes the edges) and the derived
permutation $\nu$ (which describes vertices).  Each edge
(or vertex) is equivalently described by two cycles of the
permutation $\varepsilon$ (or $\nu$).  In the algorithm we use
only one of these cycles; it does not matter which one is used.

We go through all possible assignments of values to the variables
$b_z$ such that the following conditions are satisfied:

\begin{enumerate}
\item if $(z_1\, z_2)$ is a unitary edge then $b_{z_1} \in \{i,o\}$,
  otherwise $b_{z_1} \in \{io,oi\}$.
\item the variables on the opposite sides of an edge end are related
  by
  \begin{equation}
    \label{eq:edge_end_rel}
    b_z = \wh{b}_{\wb{\varepsilon(z)} }.    
  \end{equation}
\end{enumerate}

Then, every cycle $(z_1\, z_2)$ in (the first half of) the permutation
$\varepsilon$ gives rise to the factor $E(b_{z_1}, b_{z_2})$.  Every
cycle $(z_1\, z_2\, \ldots z_k)$ in (half of) the permutation $\nu$ contributes the
factor $V_k(\wt{b_{z_1}}, \wt{b_{z_2}}, \ldots, \wt{b_{z_k}})$.

\begin{example}
  Consider the map from Fig.~\ref{fig:base_diagO}(a) in Example~\ref{exm:base_diagO}.
  The labels of the variables $b_z$ are shown in Fig.~\ref{fig:base_diag_algo}(a).
  An example assignment of the edge end, illustrated in Fig.~\ref{fig:base_diag_algo}(b), is
  \begin{equation*}
    b_1 = i, \quad b_4 = o, \quad 
    b_5 = io, \quad b_{\wb{5}} = oi, \quad
    b_2 = io, \quad b_{\wb{2}} = io,
  \end{equation*}
  with the other variables deduced using \eqref{eq:edge_end_rel}:
  \begin{equation*}
    b_3 =  \wh{b}_{\wb{\wb{2}}} = oi, \quad
    b_{\wb{4}} = \wh{b}_{\wb{\wb{1}}} = i, \quad 
    b_6 = oi, \quad 
    b_{\wb{1}} = o, \quad
    b_{\wb{3}} = oi, \quad
    b_{\wb{6}} = io.
  \end{equation*}
  Note that the edge stubs of the vertices get the opposite values.
  Altogether, the contribution of this assignment is
  \begin{equation*}
    \Eu(i,o) V_3(oi, oi, i) \Eo(io, oi) V_3(o, oi, io) \Eo(io, io).
  \end{equation*}

  \begin{figure}[t]
    \centering
    \includegraphics[scale=0.65]{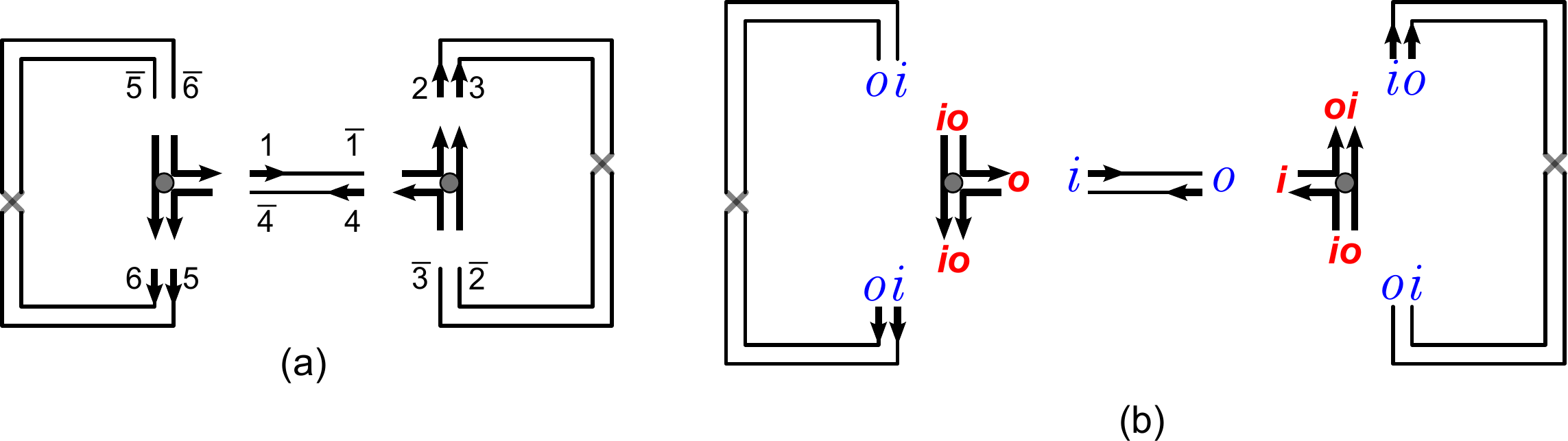}
    \caption{Edge variable labels and a possible assignment of the variables.}
   \label{fig:base_diag_algo}
  \end{figure}

  The total contribution of this base diagram is
  \begin{equation*}
    \sum_{\substack{b_1,b_4 \in\{i,o\} \\ 
        b_2,b_{\wb{3}},b_5,b_{\wb{6}} \in \{io,oi\}}}
    \Eu(b_1,b_4) V_3(\wt{b_2},\wt{b_{\wb{2}}},\wt{b_4}) 
    \Eo(b_2,b_{\wb{3}}) V_3(\wt{b_1},\wt{b_5},\wt{b_{\wb{5}}})
    \Eo(b_5,b_{\wb{6}}) ,
    \qquad b_{\wb{5}} = \wt{b_{\wb{6}}},\ 
    b_{\wb{2}} = \wt{b_{\wb{3}}} .
  \end{equation*}

  Similarly, the contribution of the base diagram of
  Fig.~\ref{fig:base_diagO}(b) is
  \begin{equation*}
    \sum_{\substack{b_1,b_3 \in\{i,o\} \\ b_2,b_{\wb4} \in\{io,oi\}}} 
    \Eu(b_1,b_3) 
    V_4(\wt{b_1},\wt{b_4},\wt{b_{\wb{1}}},\wt{b_{\wb{2}}})
    \Eo(b_2,b_{\wb{4}}) ,
    \qquad b_4 = \wt{b_2},\ 
    b_{\wb1} = b_3.
  \end{equation*}
\end{example}

As we start with a base structure with a marked half-edge and end up
with the diagram with a marked leaf, we need to account for all the
possibilities to unmark the edge and mark a leaf.  The contribution of
a diagram will be multiplied by $n/(2m)$, where $m$ is the number of
edges in the \emph{base diagram} and $2n$ is the number of leaves in
the complete diagram.  Note that our system of pre-labelling
determines which leaves are $i$ and which are $o$, so there are only
$n$ possibilities to choose the leaf $i_1$.  Since we are dealing with
generating functions with respect to $n$, the factor of $n$ will be
obtained by applying the operator
\begin{equation}
  \label{eq:factor_operator}
  s \frac{d}{ds}
\end{equation}
to the generating function of the variable $s$.

To summarize, we have sketched an algorithm to calculate the
contribution of all diagrams of a given order.  For a given order, the
number of base structures is finite.  We enumerate all of them, then
enumerate all possible leaf-markings of their edge-ends.  For each
leaf-marking we multiply together the contributions of all edges and
vertices.

\section{Moment generating functions}
\label{sec:calculation}

With the organisation of semiclassical diagrams in 
terms of principal diagrams and their untied versions and the algorithmic
approach to generate and evaluate such diagrams, we can now proceed 
to evaluate moment generating functions for various transport quantities. 

\subsection{Moments of the transmission eigenvalues}

Here we consider the typical transport problem of the linear moments of the 
transmission eigenvalues of the matrix $\bs{t}^\dagger \bs{t}$ based on the transmitting 
subblock of the scattering matrix \eqref{scatmatpartseqn} which connects the $N_1$ channels in 
one lead to the $N_2$ channels in the other.  We will obtain an 
expansion of the moment generating function
\begin{equation}
T(s) = \sum_{n=1} s^n \left\langle \Tr
  \left[\bs{t}^{\dagger}\bs{t}\right] ^n\right\rangle = NT_0(s) +
T_1(s) + N^{-1} T_2(s) + N^{-2} T_3(s) + N^{-3} T_4(s) + \ldots 
\end{equation}
in inverse powers of $N$.  

The first term
\begin{equation}
T_{0}(s)=\frac{1}{2}\sqrt{1+\frac{4\xi s}{1-s}}-\frac{1}{2} ,
\end{equation}
with $\xi=N_1 N_2/N^2$ was derived from tree recursions in
\cite{bhn08} and is valid for both symmetry classes.  The subleading
order correction requires TRS (i.e.\ $T_1^\U = 0$) and was obtained by
grafting trees onto a M\"obius strip \cite{bk11}
\begin{equation}
T_{1}^{\OO}(s)=-\frac{\xi s}{(1-s)(1-s+4\xi s)} .
\end{equation}

The next order result of \cite{bk11} could only be obtained for
reflection quantities and not for the moments of the transmission
eigenvalues.  The techniques described in the present paper allow us to
treat the transmission eigenvalues directly and to higher orders.  Much
of the semiclassical background and types of contributions were detailed
in \cite{bk11}, so we merely highlight here the results we need for the
algorithmic approach.

\subsection{Tree generating function}
\label{sec:trees_transmission}

Along with the semiclassical contributions in
Definition~\ref{def:Essen_ansatz}, we include the generating variable
$r$ with the contribution of each leaf to track the order of the
moment.  To obtain the contribution $f$ of all the unrooted trees with
a majority of $o$-leaves and $\fh$ for those with a majority of
$i$-leaves, we derive a recursive formula by cutting the trees at the
first vertex.  The trees start with an edge (contribution of $y=1/N$)
connected to a vertex of degree $2k$ (contribution $x=-N$) at which
point $2k-1$ further trees of alternating type are attached.  This
vertex can be untied if every other of these trees is an edge ending
directly in a leaf (channel).  In this case, we remove the
contributions of those edges and channels, as well as the contribution
of the vertex itself in line with Definition~\ref{def:Essen_ansatz}
while keeping the power of $r$ intact.  We therefore have the tree
recursions
\begin{equation}
f = r\zeta_2 - \sum_{k=2}^{\infty} f^k \fh^{k-1} + \zeta_2 \sum_{k=2}^{\infty} r^k\fh^{k-1} , 
\qquad \fh = r\zeta_1 - \sum_{k=2}^{\infty} \fh^k f^{k-1} + \zeta_1 \sum_{k=2}^{\infty} r^k f^{k-1} ,
\end{equation}
where $\zeta_1= N_1/N$, $\zeta_2= N_2/N$.  In the first recursion, the
first term is a tree composed of a single edge running into an
outgoing channel.  Its contribution is $yrN_2$, where $y$ is the
contribution of the edge, $r$ labels the leaf and $N_2$ counts the
number of possible choices of the outgoing channel.  In the next term,
the minus sign is the product $xy$ ($y$ being the root edge and $x$
coming from the first vertex).  Finally, in the last term, $\zeta_2$
is a product of $y$ with the $N_2$ possible choices of the one
remaining outgoing channel that every other edge is going to.  The
terms of the second recursion have similar meaning.  Performing the
sums, we have
\begin{equation}
\label{transtreessimp}
\frac{f}{1-f\fh} = \frac{r\zeta_2}{1-r\fh} ,
\qquad \frac{\fh}{1-f\fh} = \frac{r\zeta_1}{1-rf} , 
\end{equation}
which can be used to simplify the edge contributions later and which
lead to quadratic equations for $f$ and $\fh$.  However, it turns
out we will only need the generating function $h=f\fh$ which is
given by the quadratic equation
\begin{equation} \label{transheqn}
  s \xi h^2 + \left(s-2s\xi-1\right)h + s\xi = 0,
\end{equation}
where $s=r^2$ is the moment generating variable as the $n$-th moment
involves $2n$ leaves.

\subsection{Edge contributions}

To determine the edge contributions we first find the contributions of
odd and even nodes.  To create an even node (of degree $2k+2$) we
place $k$ trees of either type.  There are $(k+1)$ ways of having an
even number of trees on each side.  Such a vertex cannot be untied so
we have
\begin{equation}
yA = yx\sum_{k=1}^{\infty} (k+1)f^{k}\fh^{k} 
= - \sum_{k=1}^{\infty} (k+1) h^{k} 
= \frac{h(h-2)}{(1-h)^2} .
\end{equation}
An odd node of type $B$ also has $k$ trees of each type but they are
split to have an odd number on either side.  Such a node also cannot be
untied since the $i$ and $o$ channels are in different leads.  We get
\begin{equation}
yB = -\sum_{k=1}^{\infty} k f^{k}\fh^{k} = -\frac{h}{(1-h)^2} .
\end{equation}

The other types of odd nodes have an excess of one type of tree and so
can be untied if the alternating trees all lead directly to a channel.
For the $o$-odd node we have $(k+1)$ trees of $f$ type and the
remaining $(k-1)$ of $\fh$ type resulting in
\begin{equation}
yB_o = -\sum_{k=1}^{\infty} k f^{k+1}\fh^{k-1} + \zeta_2 \sum_{k=1}^{\infty} kr^{k+1}\fh^{k-1} 
= -\frac{f^2}{(1-f\fh)^2} + \zeta_2 \frac{r^2}{(1-r\fh)^2} = \frac{\zeta_1}{\zeta_2} \frac{f^2}{(1-h)^2} ,
\end{equation}
which simplifies following \eqref{transtreessimp} and
$\zeta_1+\zeta_2=1$.  Similarly we have
\begin{equation}
yB_i = \frac{\zeta_2}{\zeta_1}\frac{\fh^2}{(1-h)^2} ,
\end{equation}
so that the edge contributions from section~\ref{sec:algoedges} can be written as
\begin{equation}
\Eu(i,o) = \Eu(o,i) = \Eo(io,io) = \Eo(oi,oi) = \frac{(1-h)}{N(1+h)} , \qquad \Eo(io,oi) = \Eo(oi,io) = \frac{h(h-1)}{N(1+h)} ,
\end{equation}
and
\begin{equation}
\Eu(i,i) = \frac{\zeta_1 f^2(1-h)}{N\zeta_2 (1+h)}, \qquad \Eu(o,o) = \frac{\zeta_2\fh^2 (1-h)}{N \zeta_1(1+h)} .
\end{equation}
%

\subsection{Vertex contribution}

In section~\ref{sec:algovertices} we concluded that the
contribution of a vertex is
\begin{equation*}
  V_k(b_1,\ldots,b_k) = x \frac{f^q \fh^p}{(1-f\fh)^k} \chi
  = -N \frac{f^q\fh^p}{(1-h)^k} \chi,
\end{equation*}
where $\chi$ is the correction due to the possibility of untying.
Here $q$ counts how many times $i$ follows $i$ and $p$ how many times
$o$ follows $o$ in the cyclic sequence $(b_1,\ldots,b_k)$.  

For the vertex to be $i$-untied, it is necessary that $p=k$.  Each
sector then contributes
\begin{equation*}
  r + r^2 f + r^3 f^2 + \cdots = \frac{r}{1-rf},
\end{equation*}
where each $i$-tree has been substituted by a leaf, bringing $r$ to
the product.  The contribution of the untied vertex is thus
\begin{equation*}
  N_1 \left( \frac{r}{1-rf} \right)^k 
  = \frac{N_1}{\zeta_1^k} \left( \frac{\fh}{1-h} \right)^k
  = -N \frac{f^q\fh^p}{(1-h)^k} 
  \times \left( - \frac{\delta_{p,k} \delta_{q,0}}{\zeta_1^{k-1}} \right),
\end{equation*}
where the first transformation was done using (\ref{transtreessimp}).
A similar contribution comes from the $o$-untied vertex, adding up to
the total
\begin{equation}
  V_k(b_1,\ldots,b_k) = -N \frac{f^q\fh^p}{(1-h)^k}
  \left(1-\frac{\delta_{q,k}}{\zeta_2^{k-1}}-\frac{\delta_{p,k}}{\zeta_1^{k-1}}\right).
\end{equation}

\subsection{Algorithmic summation}

Plugging the above semiclassical contributions into the algorithm in
section~\ref{sec:algoapproachalgo} we can calculate the transmission
moment generating function up to order $N^{-3}$ at which point the
computational power restricts further progress.  Before listing our
answers, we consider the computation for the order $N^{-1}$ in some detail.

\begin{example}
  At order $N^{-1}$ in the absence of TRS there are only two
  contributing permutations, $\wt{\varepsilon}=(1\,4)(2\,5)(3\,6)$ and
  $\wt{\varepsilon}=(1\,3)(2\,4)$ (the corresponding maps are drawn in
  Fig.~\ref{fig:base_diagU}).  The summation over pre-labelings takes
  the form
  \begin{equation*}
    \sum_{b_1, b_2, b_3, b_4, b_5, b_6 \in\{i,o\}}
    E_u(b_1,b_4) E_u(b_2,b_5) E_u(b_3, b_6)
    V_3(\wt{b_1},\wt{b_5},\wt{b_3}) 
    V_3(\wt{b_2},\wt{b_4},\wt{b_6}),
  \end{equation*}
  and
  \begin{equation*}
    \sum_{b_1, b_2, b_3, b_4 \in\{i,o\}}
    E_u(b_1,b_3) E_u(b_2,b_4)
    V_4(\wt{b_1},\wt{b_2},\wt{b_3}, \wt{b_4}),
  \end{equation*}
  correspondingly.  We perform the summation, expressing everything in
  terms of $h = f\fh$ and $\xi = \zeta_1\zeta_2$ (note that
  $\zeta_1+\zeta_2=1$).  The answers are
  \begin{equation*}
    \frac{(2h^3\xi - 5h^3 + 4h^2\xi - 10h^2 - 6h -
      6\xi)h}{\xi(h-1)(h+1)^3}
    \qquad \mbox{and} \qquad
    \frac{-2(h^2\xi - 2h^2 + h\xi - 2h - 2\xi)h}{\xi(h-1)(h+1)^2},
  \end{equation*}
  respectively.  Including the factors $1/(2m)$ gives the total sum
  \begin{equation}
    \hat{T}_{2}^{\U}(s) = \frac{h^3(h+2)}{6\xi(h-1)(1+h)^3} -
    \frac{h^2(h+3)}{6(1+h)^3} .
  \end{equation}
  We now use \eqref{transheqn} to express $h$ in terms of $s$ and
  $\xi$, and apply the operator \eqref{eq:factor_operator} to
  arrive at the final result
  \begin{equation}
    T_{2}^{\U}(s) = s\frac{\rmd\hat{T}_{2}^{\U}(s)}{\rmd s} 
    = -\frac{\xi^2 s^2}{(1-s)^{\frac{3}{2}}(1-s+4\xi s)^{\frac{5}{2}}} . 
  \end{equation}
\end{example}

\begin{example}
At order $N^{-1}$ in the presence of TRS there are the 5 permutations $\varepsilon$ with $m=2$ edges 
and 7 permutations with $m=3$ edges listed in Table~\ref{tab:genus1_list}. 
Running through them and dividing by $2m$ we obtain the integrated moment generating function  
\begin{equation}
\hat{T}_{2}^{\OO}(s) = \frac{h^2(h^2+2h-6)}{6\xi(h-1)(1+h)^3} - \frac{h(h^2+9h-6)}{6(1+h)^3}  . 
\end{equation}
Using \eqref{transheqn} and applying operator
(\ref{eq:factor_operator}) we have
\begin{equation}
T_{2}^{\OO}(s) = s\frac{\rmd\hat{T}_{2}^{\OO}(s)}{\rmd s} 
= \frac{\xi s\left[\xi s(4s-3)+1-s^2\right]}{(1-s)^{\frac{3}{2}}(1-s+4\xi s)^{\frac{5}{2}}}  . 
\end{equation}
\end{example}

Going through the base diagrams of genus $g=3/2$ and $2$ (see Table
\ref{tab:genus_number} for their count) we can obtain the next two
orders, namely,
\begin{equation}
T_{3}^{\OO}(s)=-\frac {\xi s \left( {s}^{2}+6s+1-8{s}^{2}\xi-24s\xi+16{s}^{
2}{\xi}^{2} \right) }{ (1-s+4\xi s) ^{4}} ,
\end{equation}
with no possible permutations with broken TRS and
\begin{eqnarray}
  T_{4}^{\U}(s) &=& -{\xi}^{2}{s}^{2} (1-s)^{-\frac{5}{2}}
  (1-s+4\xi s)^{-\frac{11}{2}} 
  \Big(
  1+4s-10{s}^{2}+4{s}^{3}+{s}^{4}-20s\xi+40
  {s}^{2}\xi-12{s}^{3}\xi \nonumber\\
  & & -8{s}^{4}\xi + 9{s}^{2}{\xi}^{2} - 16{s}^{3}{\xi}^{2} +
  16{s}^{4}{\xi}^{2} \Big).
\end{eqnarray}
\begin{eqnarray}
  T_{4}^{\OO}(s)&=& \xi s (1-s)^{-\frac{5}{2}}(1-s+4\xi s)^{-\frac{11}{2}} 
  \Big( 1+20s-43{s}^{2}+43{s}^{4}-20{s}^{5}-{s}^
    {6}-99\xi s+68\xi{s}^{2} 
  \nonumber \\
  & & +326{s}^{3}\xi-448{s}^{4}\xi + 141{s}^{5}\xi + 12{s}^{6}\xi +
  518{\xi}^{2}{s}^{2} - 1304{\xi}^{2}{s}^{3} +
  1002{s}^{4}{\xi}^{2}-168{s}^{5}{\xi}^{2}
  \nonumber \\
  & & -48{s}^{6}{\xi}^{2} - 165{s}^{3}{\xi}^{3} + 408{s}^{4}{\xi}^{3}
  - 304{s}^{5}{\xi}^{3}+64{s}^{6}{\xi}^{3}\Big).
\end{eqnarray}

\begin{conjecture}
  \label{conj:transmission_form}
  From the form of $T_{2g}$ for $2g=1,2,3,4$ it is reasonable to
  conjecture that the generating functions have the general form
  \begin{equation}
    \label{eq:gen_func_conjecture}
    T_{2g}^\beta = (\xi s)^\beta (1-s)^{-(2g+1)/2} (1-s+4\xi
    s)^{-(6g-1)/2} P_{2g}^\beta(\xi, s),
  \end{equation}
  where $\beta=1$ or $2$ in the orthogonal and unitary case respectively
  and $P_{2g}^\beta(\xi, s)$ is a polynomial of order
  $2g-\beta$ in $\xi$ and $2(2g-\beta)$ in $s$.
\end{conjecture}

We note that for the unitary case $\beta=2$, we conjectured in \cite{bk11} a further grouping of
the terms in the polynomial $P_{2g}^\beta(\xi, s)$ that reduces the
number of independent coefficients.  This reduction is actually simpler in the
case of reflection coefficients which we consider below.

\subsection{Autocorrelation}

Although we have focused on the linear moments, our algorithmic
approach can be extended to non-linear statistics.  Due to the
difficulty of accounting for different tree functions $f$ and
$\fh$, we were previously unable to obtain the autocorrelation of the
transmission eigenvalues
\begin{equation}
  \tilde{P}_{[\bs{t},\bs{t}]} = 
  \sum_{n_1,n_2=1}^{\infty}s_1^{n_1}s_2^{n_2} M_{n_1,n_2}(\bs{t},\bs{t})
  - \sum_{n_1,n_2=1}^{\infty}s_1^{n_1}s_2^{n_2} M_{n_1}(\bs{t})M_{n_2}(\bs{t}) 
  = \tilde{P}_{[\bs{t},\bs{t}],1} + N^{-1}\tilde{P}_{[\bs{t},\bs{t}],2} + \ldots 
\end{equation}
beyond the first term.  For the autocorrelation, the semiclassical
diagrams have two cycles with different generating variables along
each cycle, but otherwise with trees again grafted at nodes along the
edges and at the vertices.  The result for the next term turns out to
be
\begin{eqnarray}
\tilde{P}_{[t,t],2}^{\OO}(s_1,s_2)&=&\Big[(s_1+s_2)(1-s_2)(1-s_1)^3 +8s_1^2\xi
\left(s_2^2+s_1s_2+1\right) \nonumber \\
 & & {} -2\xi\left(9s_1^2s_2+4s_1s_2^2-3s_1s_2+3s_1^3-s_2^2\right) + 8s_1^2\xi^2\left(s_1+3s_2-2s_1s_2-2s_2^2\right)\Big]\nonumber \\
 & & \times  \frac{s_1s_2}{(s_1-s_2)^{3}(1-s_1+4\xi s_1)^{2}\sqrt{1-s_2}\sqrt{1-s_2+4\xi s_2}} + \left(s_1 \leftrightarrow s_2\right),
\end{eqnarray}
where $\left(s_1 \leftrightarrow s_2\right)$ means we add the result with $s_1$ and $s_2$ swapped.
The expansion
\begin{eqnarray}
\tilde{P}_{[t,t],2}^{\OO}(s_1,s_2)&\approx& 2s_1s_2\xi(1-5\xi) + 4s_1s_2(s_1+s_2)\xi(1-9\xi+18\xi^2)  \nonumber \\
 & & {} + 8 s_1^2s_2^2\xi(1-13\xi+50\xi^2-61\xi^3) \nonumber \\
 & & {} + 6s_1s_2(s_1^2+s_2^2)\xi(1-13\xi+52\xi^2-69\xi^3) +\ldots
\end{eqnarray}
gives moments in agreement with an expansion of the results in \cite{ssw08}.

\subsection{Moments of the reflection eigenvalues}
\label{sec:calculation_reflection}

We can repeat this whole process for other transport moments, for
example the moments of the reflection eigenvalues of the $N_1\times
N_1$ matrix $\bs{r}^{\dagger}\bs{r}$ formed from the reflecting
subblock of the scattering matrix,
\begin{equation}
R(s) = \sum_{n=1} s^n \left\langle \Tr \left[\bs{r}^{\dagger}\bs{r}\right] ^n\right\rangle = NR_0(s) + R_1(s) + N^{-1} R_2(s) +\ldots 
\end{equation}
The first three terms were given in \cite{bk11} and we repeat them for
reference
\begin{equation}
  \label{eq:R_0}
  R_0^{\U,\OO}(s) = \frac{2\zeta_1 s - 1 + \sqrt{1-4\xi s}}{2(1-s)},
\end{equation}
\begin{equation}
  \label{eq:R_1}
  R_1^{\OO}(s) = \frac{\xi s}{(1-4\xi s)},
\end{equation}
\begin{equation}
  \label{eq:R_2}
  R_2^{\OO}(s) =-\frac{\xi s \left(\xi s^2 + 3\xi s - 2s +1 \right)}
  {(1-4\xi s)^{\frac52}}, \qquad
  R_2^{\U}(s) = \frac{\xi^2 s^2 (s-1)}{(1-4\xi s)^{\frac52}}.
\end{equation}
We now explain how to obtain further terms.

As the incoming and outgoing channels are in the same lead, we have
the simplification $f=\fh$ and the tree recursion reduces to
\begin{equation}
  f = r\zeta_1 - \sum_{k=2}^{\infty} f^{2k-1} 
  + \zeta_1 \sum_{k=2}^{\infty} r^kf^{k-1} , 
  \qquad \frac{f}{1-f^2} = \frac{r\zeta_1}{1-rf} ,
\end{equation}
or
\begin{equation}
  r\zeta_2 f^2 - f + r\zeta_1 =0 ,
\end{equation}
with $\zeta_2=1-\zeta_1$.

Furthermore, all odd nodes can now be untied and they all give the
same contribution
\begin{equation}
yB=yB_o=yB_i = \frac{\zeta_2 f^2}{\zeta_1(1-f^2)^2} , \qquad yA = \frac{f^2(f^2-2)}{(1-f^2)^2} ,
\end{equation} 
so that the edges provide
\begin{equation}
\Eu(i,o) = \Eu(o,i) = \Eo(io,io) = \Eo(oi,oi) = \frac{\zeta_1^2 (1-f^2)^2}{N(\zeta_1^2-\zeta_2^2f^4)},
\end{equation}
and
\begin{equation}
\Eu(i,i) = \Eu(o,o) = \Eo(io,oi) = \Eo(oi,io) = \frac{\xi f^2(1-f^2)^2}{N(\zeta_1^2-\zeta_2^2f^4)},
\end{equation}
where $\xi = \zeta_1(1-\zeta_1)$.

The vertex contribution is
\begin{equation}
V_k(b_1,\ldots,b_k) = -N \frac{f^{q+p}}{(1-f^2)^k}\left(1-\frac{\delta_{q+p,k}}{\zeta_1^{k-1}}\right),
\end{equation}
since to become untieable, the vertex has to have all $k$ sectors odd but there is no 
distinction between the $i$- and $o$-odd sectors.

As a result we obtain the following generating functions for systems
with TRS
\begin{equation}
  R_{3}^{\OO}(s) = \frac{\xi s}{\left( 1-4\xi s \right)^{4}} 
  \left(8{s}^{2}-8s+1-32s^{2}\xi+24s\xi+16s^{2}{\xi}^{2} \right),
\end{equation}
\begin{eqnarray}
  R_{4}^{\OO}(s) &=& -\frac {\xi s}{\left( 1-4\xi s \right)^{\frac{11}{2}}} 
  \Big( 1 - 26s - 52s^4\xi + 198s^4\xi^2 + 3s^6\xi^3 +
    4{s}^{5}{\xi}^{2} - 17{s}^{5}{\xi}^{3} - 87{s}^{4}{\xi}^{3}
  \nonumber \\
  & &\hspace{5mm}  {}+ 392{s}^{3}\xi -768{s}^{3}{\xi}^{2}-427{s}^{2}\xi+518{s}^{2}
  {\xi}^{2}+165{s}^{3}{\xi}^{3}-48{s}^{3} 
  +72{s}^{2}+99s\xi
  \Big).
  \nonumber
\end{eqnarray}

For systems without TRS, we also have a contribution at this last order of 
\begin{equation}
  R_{4}^{\U}(s) = 
  \frac {\xi^2 s^2 (s-1)}{(1-4\xi s) ^{\frac{11}{2}}}
  \left( 1 + 20\xi s + 9{\xi}^{2}{s}^{2} - 8s - 20\xi{s}^{2} -
    2{\xi}^{2}{s}^{3} + 8{s}^{2} - 8{s}^{3}\xi + 9{s}^{4}{\xi}^{2} \right).
\end{equation}

\begin{conjecture}
  \label{conj:reflection_form}
  From the form of $R_{2g}$ for $2g=1,2,3,4$ it is reasonable to
  conjecture that the generating functions have the general form
  \begin{equation}
    \label{eq:refl_gen_func_conjecture}
    R_{2g}^\beta = (\xi s)^\beta (s-1)^{\beta-1}
    (1-4\xi s)^{-(6g-1)/2} Q_{2g}^\beta(\xi s, s),
  \end{equation}
  where $\beta=1$ or $2$ in the orthogonal and unitary case respectively 
  and $Q_{2g}^\beta(\xi s, s)$ is a polynomial of order
  $2g-\beta$ in $\xi s$ and of order $2g-\beta$ in $s$.
\end{conjecture}

\section{Conclusions and outlook}
\label{sec:conclusions}

The algorithmic approach developed here works by creating all allowable semiclassical diagrams 
from smaller sets.  We first generate all \emph{base structure}
indexed by permutations of a certain type.  By grafting trees on the
base structures we generate \emph{principal diagrams}.  Finally we
obtain all other diagrams by untying vertices of the principal diagrams.

The base structures are organized by their genus which corresponds to
the power of $N^{-1}$ to which all the corresponding semiclassical
diagrams contribute.  For each genus, the number of base structures is
finite.  However, it grows super-exponentially with the genus and
reasonable computational limits were reached for genus 2 or the
$N^{-3}$ term in the expansion of the linear transport moments.  This
is two or three orders further than previously available results
\cite{bk11} which were obtained semiclassically and later also
recovered \cite{ms12} from an asymptotic expansion of RMT formulae
\cite{ms11}.  On the RMT side obtaining the generating functions
requires a reasonable amount of combinatorial manipulation \cite{ms12}
but these could be informed by the semiclassical results.  Similarly,
the higher order terms derived here could be useful for further
analysis of the Selberg-type integrals that appear in RMT.

Although the algorithmic approach is designed to be easily
computationally implementable, it misses the enormous scale of
cancellations that occur among the semiclassical diagrams.  For
example, from Definition~\ref{def:Essen_ansatz} each pair of diagrams
that differ by one edge and one vertex would differ by a minus sign
and cancel.  Pursuing the cancellations as in \cite{bk12,bk13a} we
could characterize the diagrams which do not immediately cancel as
primitive (palindromic) factorisations and so prove the equivalence of
RMT and semiclassics for all moments. However, this was at the cost of
making calculating moments unfeasible since such factorisations are
generated recursively like the class coefficients of RMT.  Ideally, we
would wish to include some measure of cancellation to make the
algorithm more efficient, while preserving the ease of obtaining
transport moments.  The relatively simple nature of the semiclassical
results in section~\ref{sec:calculation}, and the conjectured form for
higher terms in the $N^{-1}$ expansion, suggests that this should be
possible.

On the other hand, the fact that we are not relying on cancellation,
but generating all the diagrams, means that the algorithmic approach
will work for other physical situations where the cancellations are
not present.  For example, one might want to treat energy dependent
correlation functions which are related to Andreev billiards and the
Wigner delay times \cite{bk10,bk11,kuipersetal11,kuipersetal10}.  A
more complicated situation arises when the leads are not perfectly
coupled and a tunnel barrier exists between them and the cavity
itself.  Encounters can become partially reflected at the barriers so
that the notion of untying needs to be generalised
\cite{kuipers09,kr13,whitney07}, but one could consider extending the
algorithmic approach to cover such new possibilities.

The notion of untying is related to encounters starting or ending in
channels inside the lead \cite{wj06}.  Although a diagram and its untied version
are treated separately, classically the encounter can be continuously
moved from the lead to inside the cavity.  Governing the crossover is
the Ehrenfest time, which, when it is small compared to the average
time spent inside the cavity, separates the two cases as in
Definition~\ref{def:Essen_ansatz}.  For larger Ehrenfest time, when
RMT stops being applicable, the semiclassical treatment
correspondingly becomes notably more complicated
\cite{br06,br06b,jw06,petitjeanetal09,RahBro_prl06,wk10,waltneretal12,whitney07,wj06}.
However, a particular way of partitioning the diagrams provided enough
of a simplification that the contribution of all the leading order
diagrams could be obtained \cite{wkr11}.  This raises the possibility
that a similar partitioning could work at higher orders, and indeed
that the algorithmic approach developed here could be adapted to treat
Ehrenfest time effects.

\section*{Acknowledgements}

We are grateful to J.~Irving, K.~Richter and R.S.~Whitney for helpful
discussions.  GB is partially supported by the NSF grant DMS-0907968
and JK by the DFG through FOR 760.  We thank the anonymous referee for
many corrections and suggestions for improvement.

\appendix
\section{Target permutation of a diagram}

Here we explore the \emph{target permutation} associated
with each diagram and further formalize the notion of \emph{untying} a
vertex.

Starting with a leaf labelled $j$ on a diagram, we follow the \emph{solid} boundary
segment adjacent to it until we arrive to the leaf $\wb{j}$ from which
we follow the dashed boundary segment to the leaf number $\tau(j)$.
We can also start at the leaf $\wb{j}$, follow first the solid then
the dashed segment to arrive to the leaf number $\tau(\wb{j})$.  This
procedure defines the permutation $\tau$ which we call the
\emph{target permutation} of the diagram.  The principal diagrams, by
definition, have the target
\begin{equation}
  \label{eq:principal_target}
  \tau = (1\,2\,\ldots\, n) (\wb{n}\, \ldots \, \wb{2}\,\wb{1}),
\end{equation}
but more general diagrams are possible.  For example, untying a vertex
of the principal diagram leads, in general, to a diagram with a
different target permutation.  As defined, the permutation $\tau$ acts
on $2n$ symbols $Z = \{1, \ldots, n, \wb{n}, \ldots, \wb1\}$.  Below,
by $z$ we denote a symbol from $Z$, i.e. a label with or without the
bar.

\begin{definition}
  The \emph{reversal} of a cycle $(z_1\, z_2 \ldots z_k)$ is the cycle
  \begin{equation*}
    \wb{(z_1\, z_2 \ldots z_k)} = (\wb{z_k} \ldots \wb{z_2} \, \wb{z_1}),
  \end{equation*}
  The reversal of a permutation is performed by reversing every cycle.
  If we define the involution $\Tt(z) =\wb{z}$, which adds or removes the bar 
  with the understanding $T^2(z)=z$, then we can write $\wb{\rho} = \Tt \rho^{-1} \Tt$.
\end{definition}
 
It can be shown \cite{bk13a} that the target permutations have
palindromic symmetry: for every cycle it also contains the reversal of
this cycle (which is also required to be distinct).  Furthermore, the
targets of unitary diagrams do not mix labels with and without the bar
and can thus be thought as permutations from $S_n$: the permutation on
the symbols with bars can be recovered from the symmetry.  When we use
the $S_n$ permutation as a target in the unitary case, we call it the
\emph{reduced target permutation}.  We will now re-visit the notion of
untying, first considering the orthogonal diagrams; all consideration
can then be simply restricted to unitary diagrams.

\begin{figure}[t]
  \includegraphics[scale=0.7]{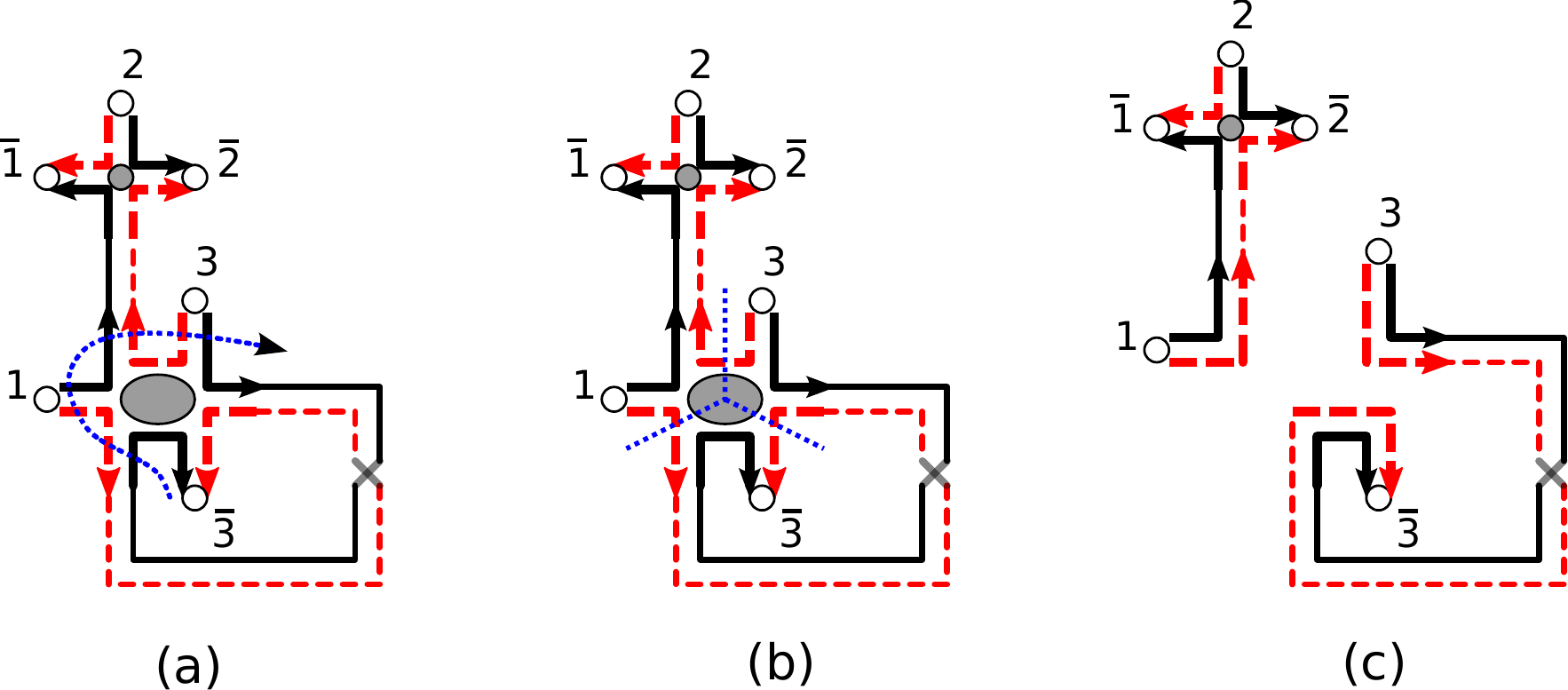}
  \caption{Untying a vertex of degree 6.  In part (a) we identify the
    6-vertex as untieable.  Its key is the cycle $(\wb{3}\, 1\, 3)$.
    To determine the direction of the key (indicated by the dotted
    line), we start with a leaf (for example, $\wb{3}$) and follow the
    solid line out of it (ignoring the direction of the solid line)
    which takes us clockwise past the vertex.  In part (b) the dotted
    line indicates how the vertex is to be cut in the untying process.
    The result of untying is shown in part (c).}
  \label{fig:untying3_TR}
\end{figure}

\begin{definition}
  \label{def:untieable}
  A vertex is \emph{untieable} if every second of
  its edges leads directly to a leaf.  The \emph{key} of the untieable
  vertex is the cycle composed out of the leaf labels read around the
  vertex.  The direction is specified by following, for a short while,
  the solid line out of one of the leaves in question (see
  Figure~\ref{fig:untying3_TR}). 
\end{definition}

An untieable vertex of degree $2m$ can be untied by cutting it into
$m$ parts, preserving the solid boundary segments.  An example of
untying is shown in Fig.~\ref{fig:untying3_TR}.  The effect of untying
on the target permutation is as follows

\begin{lemma}
  \label{lem:untying_gen_O}
  If $\tau$ is the target permutation of an orthogonal diagram, then
  after untying a vertex with key $\rho$, the target permutation
  becomes
  \begin{equation*}
    \rho^{-1} \tau \left( \wb{\rho} \right)^{-1},
  \end{equation*}
  where $\wb{\rho}$ is the reversal of the cycle $\rho$.
\end{lemma}

\begin{example}
  In Fig.~\ref{fig:untying3_TR} we have a diagram with the target
  $(1\,2\,3)(\wb{3}\,\wb{2}\,\wb{1})$ and key $(\wb{3}\, 1\, 3)$.  After untying the target is
  \begin{equation*}
    (1\,\wb{3}\,3) (1\,2\,3)(\wb{3}\,\wb{2}\,\wb{1})
    (\wb{3}\,3\,\wb{1}) = (1\,2)(3)(\wb{3})(\wb{2}\,\wb{1}),
  \end{equation*}
  which can be verified by inspecting the result in
  Fig.~\ref{fig:untying3_TR}(c). 
\end{example}

In the unitary case the definition of the untieable vertex and its key
is identical to the orthogonal case.  However, due to the symmetries
of the diagram, the key is composed of labels either all with a bar or
all without bar.  In the former case we call the vertex
\emph{$i$-untieable}, in the latter \emph{$o$-untieable}.

\begin{lemma}
  \label{lem:untying_gen_U}
  If $\tau$ is the reduced target permutation of a unitary diagram, then
  after untying a vertex with key $\rho$, the reduced target permutation
  becomes
  \begin{equation*}
    \begin{cases}
      \rho^{-1} \tau, &\mbox{$i$-untying},\\
      \tau \wb{\rho}^{-1},&\mbox{$o$-untying},
    \end{cases}
  \end{equation*}
  where $\wb{\rho}$ is the reversal of the cycle $\rho$.
\end{lemma}

\begin{remark}
 \label{remark:unitsubsetorth}
 Lemma~\ref{lem:untying_gen_U} is obtained from
 Lemma~\ref{lem:untying_gen_O} by simply erasing all cycles made of
 symbols with bars.  One can go in the other direction as well,
 reconstructing the cycles with bars using the palindromic symmetry.
\end{remark}

 The above lemma further explains the notation we used for the untied
versions of the principal diagrams in (\ref{eq:M2}).

\begin{figure}[t]
  \includegraphics[scale=0.7]{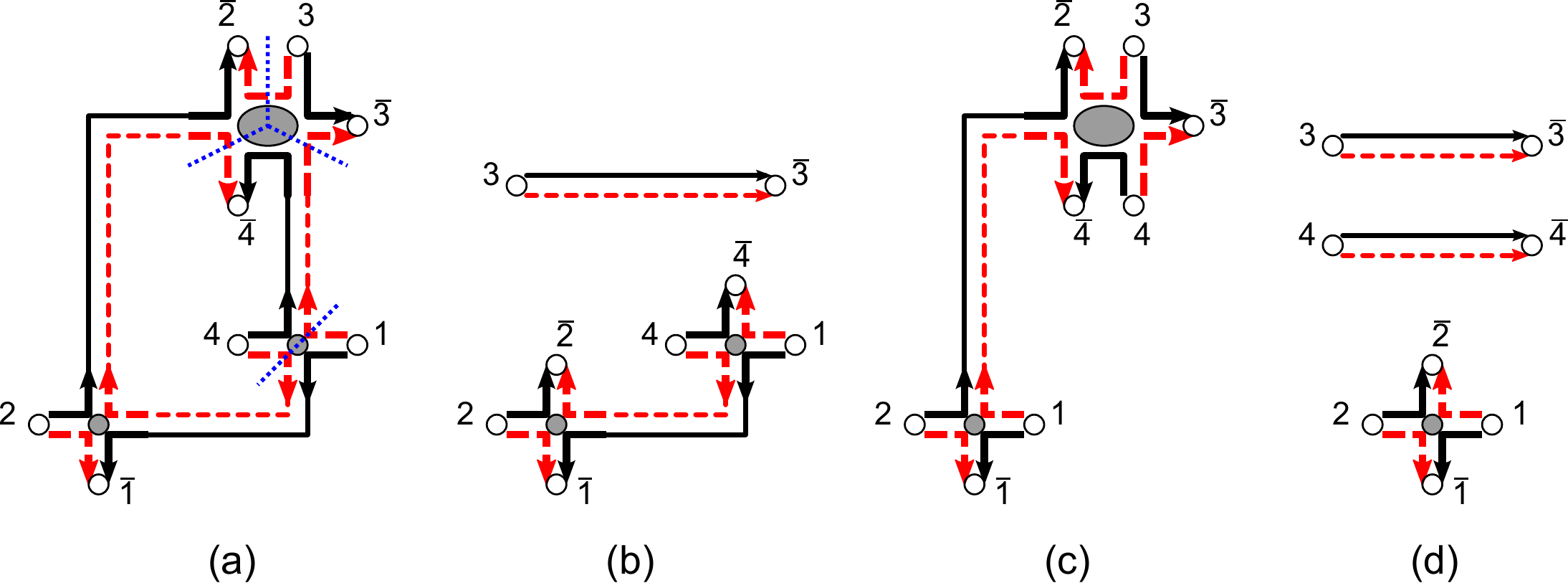}
  \caption{(a) A unitary diagram with two untieable vertices.  Untying
    the 6-vertex leads to (b) while untying the 4-vertex leads to
    (c). Untying both results in diagram (d).}
  \label{fig:untying3_NTR}
\end{figure}

\begin{example}
  Consider the diagram in Fig.~\ref{fig:untying3_NTR}(a).  Its reduced target
  permutation is $(1\,2\,3)(4)$.  The vertex of degree 6 is
  $o$-untieable with the key $(\wb{2}\,\wb{4}\,\wb{3})$.  After
  untying, the reduced target of the diagram becomes $(1\,2\,3)(4)(2\,4\,3) =
  (1\,2\,4)(3)$ as in Fig.~\ref{fig:untying3_NTR}(b).  The vertex of degree 4 is $i$-untieable with the key
  $(1\,4)$.  The reduced target after untying is $(1\,4) (1\,2\,3)(4) =
  (1\,2\,3\,4)$ depicted in Fig.~\ref{fig:untying3_NTR}(c).  Both can be untied 
  giving $(1\,4)(1\,2\,3)(4)(2\,4\,3) = (1\,2)(3)(4)$ as in in Fig.~\ref{fig:untying3_NTR}(d).
\end{example}

\bibliographystyle{abbrv}
\bibliography{ctsetm}

\end{document}